\documentclass[preprint2]{aastex}

\input psfig

\def\ie{\hbox{\it i.e.,\ }}
\def\eg{\hbox{\it e.g.,\ }}
\def\deg{$^\circ$}

\def\etal{\ et~al.}
\def\gs{\mathrel{\lower0.6ex\hbox{$\buildrel {\textstyle >}
 \over {\scriptstyle \sim}$}}}
\def\ls{\mathrel{\lower0.6ex\hbox{$\buildrel {\textstyle <}
 \over {\scriptstyle \sim}$}}}
\def\micron{\hbox{$\mu\rm m$}}

\def\Halpha{\hbox{$\rm H\alpha$}}

\def\Hbeta{\hbox{$\rm H\beta$}}

\def\deg{\hbox{$^{\circ}$}}

\def\e#1{\times 10^{#1}} 

\begin{document}

\title{Microslit Nod-shuffle Spectroscopy --- a technique for achieving
very high densities of spectra
}

\author{Karl Glazebrook\footnote{Present address:
Department of Physics \& Astronomy,
Johns Hopkins University,
3400 North Charles Street,
Baltimore ,
MD 21218-2686, 
USA. {\it kgb@pha.jhu.edu}} \&
Joss Bland-Hawthorn}
\affil{Anglo-Australian Observatory, \\ PO Box 296, Epping, NSW 1710, \\ AUSTRALIA
}

\begin{abstract}
We describe a new approach to obtaining very high surface densities of
optical spectra in astronomical observations with extremely accurate subtraction
of night sky emission. The observing technique requires that the telescope is 
nodded rapidly between targets and adjacent sky positions; object and sky spectra 
are recorded on adjacent regions of a low-noise CCD through charge shuffling.
This permits the use of extremely high densities of small slit apertures 
(`microslits') since an extended slit is not required for sky interpolation. 
The overall multi-object advantage of this technique is as large as 2.9$\times$ 
that of conventional multi-slit observing for an instrument configuration 
which has an underfilled CCD detector and is always $>1.5$ for high target 
densities. The `nod-shuffle' technique has been practically implemented at the 
Anglo-Australian Telescope as the ``LDSS++ project'' and achieves sky-subtraction 
accuracies as good as 0.04\%, with even better performance possible.  This is a factor
of ten better than is routinely achieved with long-slits.
LDSS++ has been used in various observational modes, which we describe, 
and for a wide variety of astronomical projects. The nod-shuffle approach should be of 
great benefit to most spectroscopic (\eg long-slit, fiber, integral field) methods
and would allow much deeper spectroscopy on very large telescopes (10m or greater) than
is currently possible. 
Finally we discuss the prospects of using nod-shuffle to pursue extremely long spectroscopic
exposures (many days) and of mimicking nod-shuffle observations with infrared arrays.
\begin{center}
{\it Accepted for publication in PASP}
\end{center}
\end{abstract}

\keywords{instrumentation: detectors --- instrumentation: spectrographs 
--- techniques: spectroscopic}

\section{Introduction}
The problem of subtracting the night sky foreground emission is
a critical one for astronomical spectroscopy. The task is
particularly acute in the red part of the spectrum (600$-$1000nm)
as there are numerous hydroxyl (OH) bands which dominate the light
giving a bright background. Many authors have recognized over
the past twenty years that low to moderate resolution spectroscopy 
in this band is ultimately limited by systematic uncertainty 
associated with sky subtraction (\eg Dressler 1984).

In some respects, it is surprising that optical astronomy has been 
slow to recognize an important technique utilized by near-infrared 
astronomy, \ie beam-switching. Here, the background signal is very
strong, is highly variable, and influences all observations (\eg 
Ramsay, \etal, 1992). A common implementation of 
beam-switching is where the secondary mirror `chops' between a 
target object and a sky field while the infrared array is read out 
continually.\footnote{`Chopping' refers to a moving secondary
mirror while the primary remains fixed on the object; we use `nodding' 
to indicate a fixed secondary where the pointing of the primary 
mirror alternates between sky and an object field.}

This is perhaps because there is a conflict 
between the desire to beam-switch rapidly, and
sample the sky contemporaenously, and the desire to take long integrations
to minimize the effect of readout noise. This is especially true for modern, very
low noise CCD detectors. 

The underlying principle of the nod-shuffle technique is simply that a CCD
detector can be used to store two images of a field, imaged quasi-simultaneously
(Cuillandre\etal\ 1994; Bland-Hawthorn 1994; Sembach \& Tonry 1996). By using
`charge-shuffling' charge can be moved from an  illuminated
region to a storage region. This process does not invoke readout noise and
only takes only a fraction of a second
since charge can be shifted between CCD rows two to three orders of
magnitude faster than it can be read out. If this shuffling is
synchronized to telescope motion two interleaved exposures of object and 
sky can be imaged side by side at the detector. 
Note three important facts: (i) the images 
are obtained through identical optical paths, (ii) the imposed flatfield
structure is identical for both images, and (iii) the CCD is read out 
only once.  

The use of shuffling techniques in astronomy can be traced to 
early attempts to improve the performance of imaging polarimeters 
(McLean\etal\ 1981; Stockman 1982).  Since that time, charge-shuffling 
has been little utilised.  Part of the reason may stem
from experiments by Lemonier \& Piaget (1983). By rapidly shifting 
charge backwards and forwards many times (pocket pumping), they were able to 
identify local defects in the potential profile (trapping sites)
within the silicon substrate. By the end of the 1980s, traps and
deferred charge were still a fundamental limitation to repeated 
charge shuffle operations (Blouke\etal\ 1988).

The development of charge-shuffling at the Anglo-Australian Observatory
dates back to the 1994 Marseilles conference 
on imaging spectrographs (Comte \& Marcelin 1995).  It was here that 
the first results of
integral field spectrographs were presented, arguably the most
important development in optical instrumentation in the past decade.
It was clear, and remains true, that the fundamental limitation of 
this powerful technology is the difficulty of accurate sky subtraction
(Bland-Hawthorn 1995).

Key developments in CCD manufacture have made charge-shuffling a
realistic prospect and an important 
consideration in all future instrument design. First, 
the latest generation CCDs (EEV, MIT Lincoln Lab) have very low 
read noise ($\sim 1e^-$), negligible dark current, high purity and very 
high charge transfer efficiency (99.9999\%). Secondly, the manufacturing
process prefers to generate rectangular arrays\footnote{The photofab process uses a 
$25-30$mm reticle which restricts the `row' dimension of a CCD. The 
reticle is stepped down the wafer and the new circuit is stitched to 
the previous pattern.} which provide for
storage regions. 
Bland-Hawthorn \& Barton (1995) demonstrate that, with modern 
CCDs, it takes more than a hundred shuffle operations before bulk trapping 
sites start to compromise the data. 
 
In this paper, we describe the development at the AAO of the `nod-shuffle' 
method founded on the principle of CCD charge shuffling.  This differential 
technique has resulted in two important experimental breakthroughs. First, 
the object and sky can be measured quasi-simultaneously. As we show, the main 
limit to the accuracy of sky-subtraction is the rapidity of nod-shuffling
compared to the temporal power spectrum of sky brightness variations.
Secondly, nod-shuffle allows for a considerable increase in the 
{\it multi-object gain} of a spectrograph,  up to 2.9$\times$ more objects per 
unit observing time using small `microslits,' for fields with high object
densities. We have implemented the nod-shuffle method with the Low Dispersion
Survey Spectrograph (LDSS) on the Anglo-Australian Telescope (AAT) and have
obtained fractional residuals as low as $4\e{-4}$.

The plan of this paper is as follows: in Section~2 we describe the nod-shuffle
concept and discuss qualitatively the sky-subtraction and multiplex advantages 
to be gained. In Section~3 we describe in detail our implementation of 
nod-shuffle at the AAT using the Low Dispersion Survey Spectrograph and show 
some example data. In Section~4, we show the increased multi-object gain which 
becomes possible via the nod-shuffle operation. We quantify the sky-subtraction 
accuracy in Section~5 and discuss ways in which it might be improved further.  In 
Section~6, we illustrate key observing modes for LDSS++ which are facilitated by
the use of microslits.  Finally, we discuss future prospects for the 
nod-shuffle observing mode.

\section{The Nod-Shuffle Concept} \label{sec-concept}

The concept behind charge shuffling is that unilluminated portions of
a CCD can be used for storage. The image formed on an illuminated
portion can be `shuffled' very quickly into a storage area by clocking 
before being shuffled back at a later stage. For example, with the AAO-1 CCD
controller and the Thompson 1024$\times$1024 format CCD, a single row can be 
shifted upwards or downwards in 12.5~$\mu$s, compared to 30--160~ms when 
clocked through the output amplifiers.\footnote{The AAO-1 controller was
upgraded in 1998 resulting in a fivefold increase in pixel rate. But this is
still three orders of magnitude slower than the rate that charge can be
shifted between rows without reading out.} The shift operation is a factor 
of 4 slower for the Tek 1024$\times$1024 format and MITLL 2048$\times$4096 
format CCDs. Since the shuffle operation does not involve the read-out 
amplifiers, the primary source of noise is now associated with charge 
transfer within the substrate (Janesick \& Elliott 1992).

Each vertical clock shifts the complete image on the CCD one row up
towards the readout register. The row that was next to the readout
register gets clocked in to the readout register and cannot be reverse
clocked back into the image.  At the other end of the image, a `clean'
row is generated. This happens for shifting in the `forward' direction.
Clocking in the reverse direction moves the complete image one row
away from the readout register for every vertical clock applied to the CCD. 
A clean row is generated next to the readout register and at the
other end one image row is lost.  

\begin{figure*} 
\psfig{figure=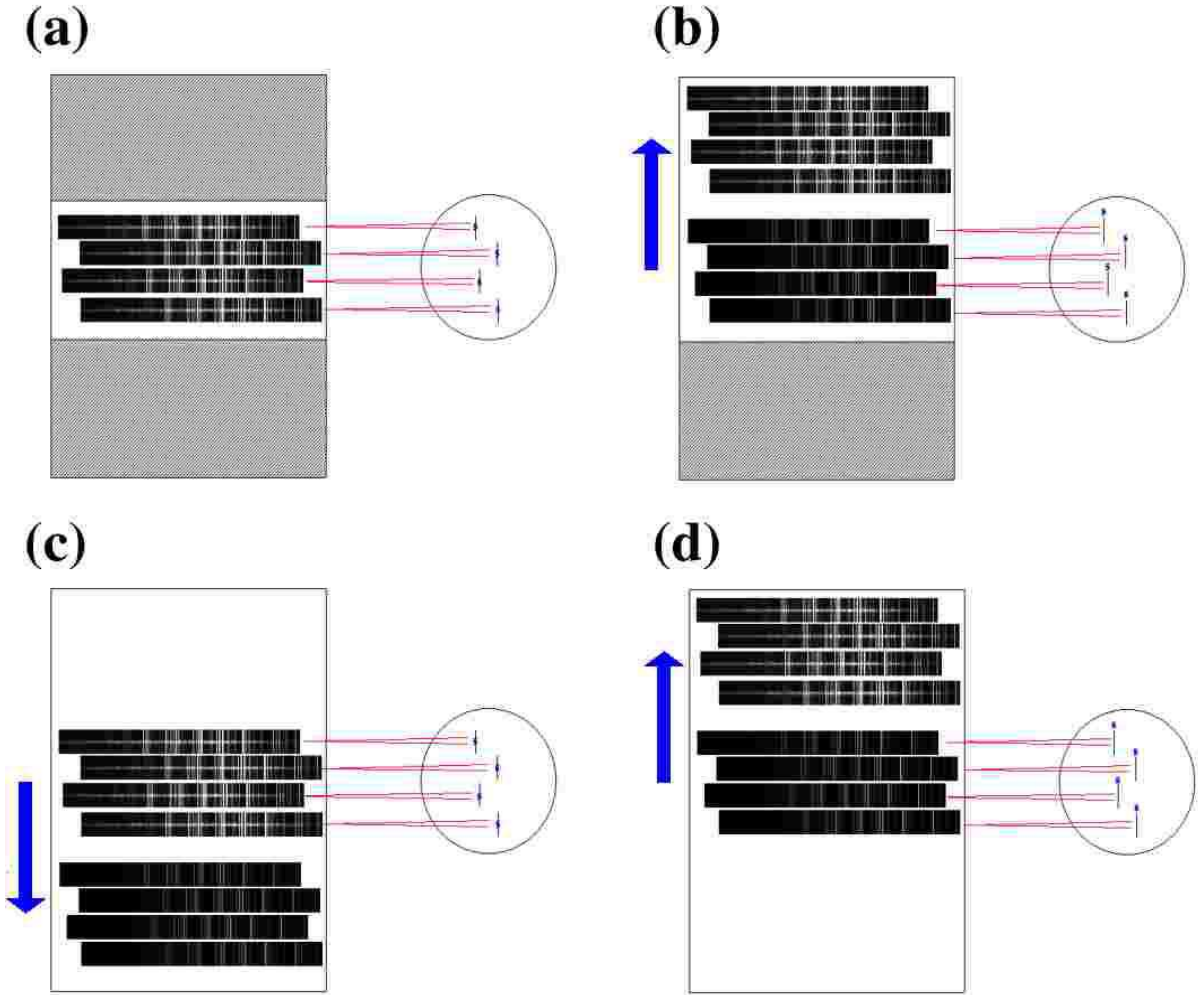,angle=-90,width=\hsize}
\caption{Illustration of the nod-shuffle procedure implemented
in the LDSS spectrograph showing progressive
stages of image formation: (a) The spectra of the objects through the
slits is imaged onto the central portion of an oversized CCD. (b) The first
image is shuffled up into a storage region (with the shutter closed), the 
telescope is offset to adjacent sky which is then imaged onto the now
empty central region of the detector. (c) The object image is shuffled back
and additional object photons are imaged (d) Sky is shuffled back and 
imaged. Steps (c) and (d) are cycled continuously until the integration
is complete.}\label{fig-nodshuffle}
\end{figure*}

In order to produce two contiguous images side by side 
on the detector via shuffling, the maximum field of view (\ie number of rows) 
which can be shifted without loss of information for the exposed or 
stored image is one third of the detector's column dimension. The 
reason is clear: when the detector is
clocked in one direction, rows at the detector edge are lost
(c.f. Figure~\ref{fig-nodshuffle}).
More generally, shuffling between $m$ partitions uses
$m/(2m-1)$ of the CCD for holding the separate observations,
while the remainder is (a) used for temporary storage, and (b) rendered 
useless by the shuffle process (\ie this fraction is never illuminated).
A fuller technical description of charge-shuffling is given in 
Bland-Hawthorn \etal\ (2000). 

The nod-shuffle image sequence developed for LDSS observing
is utilises this underfilled, large-shuffle mode and is 
illustrated in Figure~\ref{fig-nodshuffle}.
The observing sequence is as follows:

\begin{enumerate}
\item The target objects are acquired with the telescope on to the spectrograph mask slits
(these may be true slits or simple apertures such as holes).
\item The shutter is opened for a OBJECT exposure (usually 10--100 secs in duration), 
dispersed spectra of OBJECTS$+$SKY are accumulated in the central area.
\item The shutter is closed.
\item The OBJECT image is shuffled up, by clocking the CCD charge pattern. to a
upper storage area which is unilluminated.
\item The telescope is moved to a SKY position. (This can be a truly blank patch or
can simply involve moving the objects some way along the slits).
\item The shutter is re-opened and dispersed SKY spectra are accumulated, for the
same exposure time as the OBJECT, in the blank central area.
\item The shutter is closed, the charge is shuffled back down bring the OBJECT image
back in to the center and the SKY image into blank storage. The telescope is moved
back to the OBJECT position.
\item The shutter is opened and more OBJECT data is accumulated.
\item The sequence OBJECT--SKY--OBJECT--SKY--... is repeated for the rest of
the exposure.
\end{enumerate}

At the AAT, the OBJECT and SKY exposures are typically 30 secs, repeated to
fill up a 1800 sec exposure before readout. Sky subtraction then consists
of extracting the two regions and calculating the difference image. This
technique, which we call ``nod-shuffle,'' gives extremely precise sky-subtraction
for the following reasons:

\begin{description}
\item[a] The OBJECT and SKY are observed through identical slits/apertures. The effect of
any irregularities cancel out in the subtraction.
\item[b] The OBJECT and SKY are imaged on to the exactly the same pixels on the detector.
The optical path is identical. The pixel response is identical. (The response is that
of the pixel where the image is {\em measured} --- the storage pixels have no effect).
\item[c] The OBJECT and SKY are observed quasi-simultaneously, thus the effect
short timescale temporal sky variations cancel out in the subtraction. This is
quantified below in Section~\ref{sec-sky}.
\item[d] The OBJECT and SKY positions can be extremely close (a few arcsecs) so spatial
sky variations are not significant.
\item[e] Because of the identical light path and quasi-simultaneity the effects of
fringeing on the detector from night sky lines cancels out.
\item[f] Similarly the effects of any instrument flexure during the course of the
exposure cancel out.
\item[g] There is no need to re-sample and interpolate the sky for the subtraction,
so there are no numerical artifacts introduced.
\item[h] The presence of any DC level in the detector due to bias, dark current,
or scattered light does not affect the sky-subtraction. If it is constant it
cancels, if it varies across the detector (including the unilluminated regions)
it will not cancel but will still not affect the sky subtraction. 
\end{description}

\begin{figure}
\psfig{figure=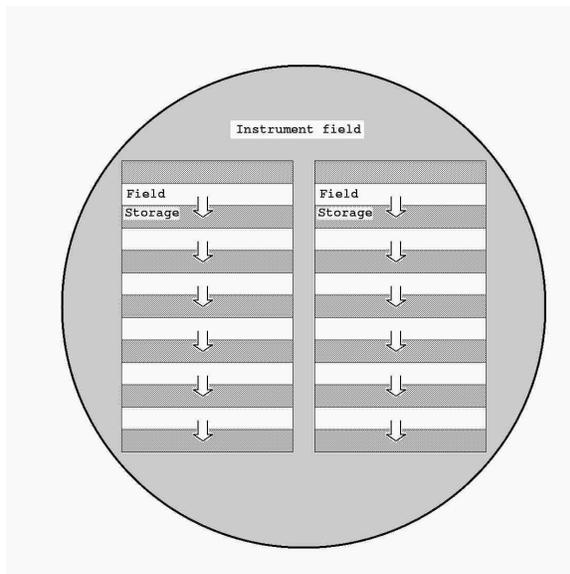,width=\hsize}
\caption{Illustration of the nod-shuffle geometry in the case
in which the detectors are overfilled by the instrument field of view (FOV)
(in this case two detectors are shown). Unilluminated regions must
be taken from the active FOV giving a 50\% overhead resulting in a
stripe pattern. Note the stripe width can be as small as individual 
spectra, however it is desirable to make them larger to minimise
area lost to edge effects at the strip boundaries: a region of wide
$\simeq$ instrument PSF will be badly subtracted. A reasonable width
would be large banks 20--50 spectra.} \label{fig-overfill}
\end{figure}

Of course this is a much more complex observing sequence than simply acquiring 
objects on to slits and staring. There is also a penalty for the precise 
sky-subtraction: $\sqrt{2}$ more noise in the resulting spectra because
of the subtraction, compared to a very long slit, though the systematics
in the sky removal are expected to be greatly improved.

However nod-shuffle offers another great advantage over conventional multislit
spectroscopy: it permits a large increase in the achievable object multiplex.
Because a long slit is no longer required for sky-subtraction via interpolation the
apertures only need be large enough to cover the objects. We term these
``microslits''. Additionally they need not be slits --- they can be apertures of any shape
such as circles. If we take the example of observing faint 24$^{\rm th}$ magnitude
galaxies only a 1 arcsec aperture is required due to their small size (Smail\etal\ 1995). 
Comparing this to typical multislit observations with 10--15 arcsec long slits 
(Glazebrook\etal\ 1995), we can see that we would expect 10--15$\times$ as many slits 
to be squeezed on to the
mask without spectral overlap. We quantify these 
multiplex gains below in Section~\ref{sec-mplex}.

Finally we note that for multi-object spectroscopy there is an alternative
mode of observing where the charge is shuffled only a few pixels. Because
a slit mask blocks out light any part of the CCD can be used for storage.
This is particularly useful because it scales to multiple, mosaicked CCDs
i.e. when the camera FOV is much bigger than
the detectors. This case is illustrated in Figure~\ref{fig-overfill}. A penalty here is that
half the available detector area must be used for storage when it could be used
on-sky, however as we demonstrate below it still gives a formal multiplex advantage
in the high source density limit. 

\section{The AAT/LDSS++ Implementation} \label{implementation}

The practical implementation we will describe was developed using the AAT's
Low Dispersion Survey Spectrograph (LDSS), which came to be known as the
LDSS++ project. LDSS is a wide-field multislit spectrograph with a 12 arcmin
field of view. A large collimator re-images the telescope pupil, in this
space can be inserted grisms and/or filters, this is then imaged through a
camera onto a CCD detector (Wynne \& Worswick 1988; Glazebrook 2000).
The grism can be taken out for direct imaging of the field or the mask, 
this is used to acquire the field on to the mask accurately.

LDSS has recently been equipped with a volume-phase holographic grating (VPH; Barden, 
Arns \& Colburn 1998) and a MITLL deep-depletion $2048\times 4096$ CCD detector with 
15\micron\ pixels. These two upgrades give a considerable improvement in
the red 500--1000nm throughput of the system: the gain at 700nm is a factor
of 2 (Glazebrook 1998).

The LDSS field of view is circular and is $\simeq 2000$ pixels on the detector
(0.39 arcsec pix$^{-1}$ scale). The shuffle direction is along the
long axis of the CCD, perpendicular to the dispersion direction exactly as shown in
Figure~\ref{fig-nodshuffle}. This is not absolutely necessary but is 
done because it is easier to block the adjacent storage areas spatially by using
the mask; otherwise some sort of spectral blocker would be required and this would
not be ideal due to offsets between slits. In nod-shuffle mode we thus use the
central $2048\times 1365$ pixels. It represents approximately 
the underfilled case described
in Section~\ref{sec-concept}.

The implementation of our nod-shuffle scheme is as follows.
At the start of a nod-shuffle run, a shuffle sequence is
downloaded to the CCD controller micro and the instrument sequencer
micro from the VAX computer; the instrument sequencer also receives
a telescope command set.  The VAX then tells the instrument sequencer 
and the CCD controller to `run'.  The controller runs software
which interprets the shuffle sequence, clocking the charge up and down
and driving the CCD shutter. It 
dictates each step by triggering an event with an `external sync' 
pulse for each phase of the operation.  The triggers occur after fixed 
time intervals since there is presently no handshake from the telescope.  
The number and nature of the triggers depend on whether there is to be
guiding at either the object or sky position (OFFSET mode), at neither 
(OFFSET NO GUIDE mode) or both (AXES mode).
With the output pulse, the CCD controller toggles the status of 
an I/O line and waits for a given delay time. The instrument sequencer 
reads the I/O line and, when required, writes telescope control commands 
to a port on the VAX/VMS computer system. A program running on the
VAX reads these commands, translates them and routes them via the CAMAC
interface to the telescope control Interdata computer.

There is no feedback in this system: the CCD controller does not know
the state of the telescope. Ideally of course it would, but this would 
require complete re-engineering of the whole observing system. Instead 
the telescope movement is allowed for by predetermined time delays.
The controller waits a given amount of time between shuffles
with the shutter closed to allow the telescope to finish its `offset
and stabilise' action. For small offsets of a few arcsec, the AAT does 
this in about 1 second; typically we allow 2 secs dead time in a 30
sec integration time. It was 
verified that this was adequate by taking long-exposure direct images 
of star fields in nod-shuffle mode and
looking for image elongation along the offset direction. The two
shuffled images can also be subtracted to look for elongated residuals
--- none were found down to the noise level.

\begin{figure*} 
\vspace*{-0.5in}
\psfig{figure=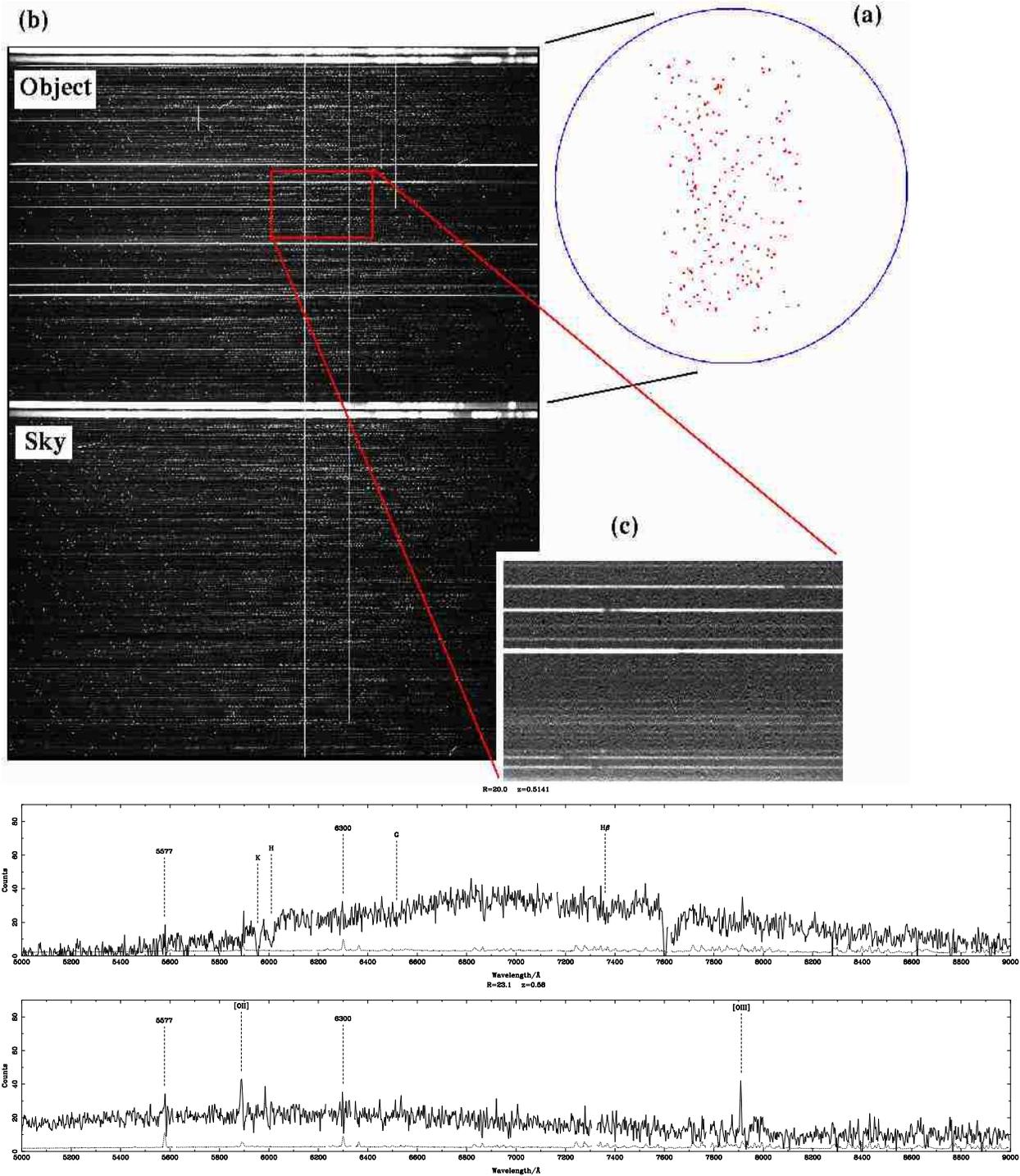,angle=-90,height=5in}
\psfig{figure=fig3b.ps,angle=-90,width=\hsize}
\psfig{figure=fig3c.ps,angle=-90,width=\hsize}
\caption{ Sample data from the HDF-S observing campaign. Panel (a) shows the
slitmask used (225 microslits) and panel (b) shows the raw shuffled data. (c) shows
the difference image zoom,ed in. The slits in the case are circular apertures, so the
spectra appears as tramlines a few pixels wide horizontally across the detector. 
Panel (d) shows two sample extracted spectra of a bright and faint
galaxy, the solid lines are the spectra
(unfluxed) and the lower dotted lines show the theorectically achievable
noise level as determined by shot and read noise (shot dominates). Bad columns are
masked out of the plot. Sky residuals are only
seen near extremely bright lines (5577\AA\ and 6300\AA\ are marked as examples)
and are entirely consistent with pure Poisson variance.
}\label{fig-sample}
\end{figure*}

Some sample data of nod-shuffle spectra are shown in Figure~\ref{fig-sample}.
This was taken for a redshift campaign in the Hubble Deep Field South 
(Glazebrook\etal\ 2000a) during commissioning of the
nod-shuffle system. 
We placed 225 microslits (circular $\simeq 1$ arcsec apertures)
on targets along the 1365 pixel spatial axis ($\equiv$ 9 arcmin),
the spectra are dispersed along the horizontal 2048 pixel axis ($\equiv 5300$\AA). 
The LDSS PSF is a Gaussian with 2 pixels FWHM at the field field 
degrading to 3 pixels at the field edge. The microslits are spaced at intervals of at 
least 4 pixels vertically (subject to target availability) so their spectra are 
significantly separated.  The horizontal spread of the slits was up to 3 arcmin so as 
not to introduce significant wavelength offsets between spectra.

It can be seen in Figure~\ref{fig-sample} that the form of this data is somewhat akin to
spectra from fiber optic spectrographs in that each object produces a tramline which is
traced and extracted. However in this case the extraction is
done {\em after} sky-subtraction and there are significant wavelength offsets between
spectra.


\section{Multiplex gains} \label{sec-mplex}

To quantify the multiplex gain we must compare the number of
spectra observable per unit time to the same limiting signal$/$noise
ratio versus the longslit case where the sky is subtracted
by interpolation. We must observe for longer with
nod-shuffling to reach the same signal$/$noise, however this is more than
balanced by an increase in the number of slits we can fit on the
mask. We call this the `nod-shuffle advantage' (NSA).

The OBJECT$-$SKY subtraction in the nod-shuffle case introduces 
$\sqrt{2}$ extra subtraction noise. First we consider at what length the longslit
subtraction introduces the same amount.  We will 
assume the longslit has length $n$ elements,
where an element is taken as the spatial extent of the target objects
(thus $n=1$ for the microslit).

Conventionally the background
along the slit, excluding the object, is fitted with either a linear model
or a higher-order polynomial. Typically the background level will vary by
a few percent across the slit due to instrumental effects such as slit
alignment and optical distortion and this slope will vary with wavelength 
due to the structure in the sky spectrum. The fitting will also be
limited by the presence of slit irregularities. This is discussed in more detail 
below in Section~\ref{sec-sky}. For now we will compute the ideal
limit for a smooth slit.

Accurate sky-subtraction in the neighbourhood
of the bright sky emission lines requires fitting at least a general linear
model to each wavelength channel, thus the error at the object location
is the error on the intercept on the slope ($\sigma_c^2$) 
from the line fitting from $N$ points:
$$ \sigma_c^2 = {2 \sigma^2 (2n+1) \over n(n-1)} \simeq {4 \sigma^2\over n} \ \ \ (n\rightarrow\infty)$$
where $\sigma$ is the noise on each point and for simplicity we have ignored
the omission of the central object point. We now consider the following question:
as the slit length $n$ increases at what point does subtracting the linear fit introduce 
less noise than nod-shuffling, i.e. $\sigma^2+\sigma_c^2<2\sigma^2$? 

This occurs at $n=6$, after allowing for more complex formulae where the central point
is omitted.  Instead of the slit we could in principle substitute 6 microslits. 
There is a factor of two nod-shuffling overhead either temporally
(due to the sky position) or spatially (if we move the object between adjacent slit
positions we have 3 pairs rather than 6 objects). 
For $n=6$ we calculate $\sigma_c^2=0.95 \sigma^2$ and so the NSA is calculated to be 2.9.
As slits become longer the NSA increases further and tends to $n/4$
for large $n$. While we can fit on $n\times$ more slits we have to observe
$4\times$ longer to allow for the two positions overhead and subtraction noise. 

In practice, as the slit becomes longer more instrumental effects come into play
and a linear fit no longer improves the residuals.  Often a higher-order polynomial is
used to allow for curvature, however this will introduce yet more noise as there
are more free parameters. In practice a slit length of 15--20 arcsec is the useful
limit, if slits are this large the NSA is 4--5. 

For the overfilled case illustrated in Figure~\ref{fig-overfill} there is another
additional factor of two for charge storage in regions which could otherwise
be used for observations; nevertheless the NSA is still 1.5 exceeding the
longslit case and providing better sky-subtraction.

Of course the theoretical NSA is only achieved if the object density is high
enough to allow close spacing of microslits. In the very low density regime where a
very long slit can be placed on each object with no concomittant multiplex loss
the NSA is only 0.5, i.e. we must
observe twice as long to balance the $\sqrt{2}$ subtraction noise. In
practice however for faint spectroscopy typical slit spectroscopy {\em is} dominated
by residual systematics at the 0.5--1\% level (see Section~\ref{sec-sky}) and
not random noise where the lines are bright. And at low resolutions ($R<2000$) a
large fraction ($\sim 50\%$) of the red spectrum is occluded by bright lines,
so the supposed $S/N$ loss is moot.

One common technique to reduce these sky residuals in otherwise conventional
longslit observing is to use a `slow' beam-switching
technique to improve the systematic residuals when observing ultra-faint targets
by moving the object along the slit in consecutive observations. This is analogous
to nod-shuffle except the CCD is read out between the two positions. The individual
exposures must be at least 5--10 minutes (on a 4m telescope) 
to obtain enough sky signal to be background
limited and consequently when the images are subtracted
there is a residual due to temporal sky changes. This residual is removed again by
fitting along the slit, but the systematics are reduced because of the lower overall level.
Like nod-shuffle this will always introduce $\sqrt{2}$ more subtraction noise. The minimum
NSA versus this case is now 5.9 (underfilled) and 2.9 (overfilled).

So far we have made the assumption that an independent linear fit must be done
for each wavelength. However if the sky background has no structure, i.e. is observed
in a wavelength region of featureless continuum, then we would expect the slope
across the slit to vary only slowly with wavelength and the fitting can in
principle be highly constrained. The underfilled NSA reaches 0.5 in this limit. 
However even in the blue part of the optical spectrum (350--500nm) there is still
considerable stucture in the night sky spectrum due to scattered solar absorption
lines.

Finally the NSA is maximised at very high target densities. The required
density is approximately:
$$ \rho = {3600 \beta \over W \alpha \, x} \ \ \ \hbox{objects\ arcmin}^{-2}$$
where $\beta$ is the dispersion in \AA$/$pixel, $\alpha$ is the spatial scale in 
arcsec$/$pixel, $x$ is the microslit size in arcsec and $W$ (in \AA) is either 
the wavelength range on the detector (when the spectra are short compared to the detector
size) or the minimum wavelength overlap required for all objects by the mask design
(when the spectra are comparable to or longer than the detector). 
For LDSS++, $\alpha=0.39$ arcsec pix$^{-1}$, $\beta=2.6$ \AA\ pix$^{-1}$, for the HDF-S project we 
used $W=3000$\,\AA  and $1.0$ arcsec apertures. This gives a sky density requirement of
$\simeq 8$ objects arcmin$^{-2}$. For field galaxies this density is achieved at 
$R\approx 23$ (Hogg\etal\ 1997; Smail \etal\ 1995). It is also very suitable 
for observing stellar and galaxy clusters. It is a much higher density than can be achieved
by conventional multislits ($\sim 5$--$10\times$) and by  fiber spectrographs --- for example
the highly multiplexed 2dF spectrograph can only reach 0.05--0.1 objects arcmin$^{-2}$
(Lewis\etal\ 2000).

\section{Sky subtraction accuracy} \label{sec-sky}


\subsection{Achievable accuracy with conventional multi-slits } \label{sec-sky-mslit}

In order for the figures for nod-shuffle accuracy to be meaningful, it is
useful to consider how well sky can be subtracted using a longslit. This
is limited by instrumental imperfections such as variable PSF, slit and CCD
irregularities, slit tilt and pixel sampling effects, image distortion, fringeing, flexure
etc. The effect of slit tilt, with respect to the CCD columns, is particularly intresting
as it is this which causes linear sky variations across the slit. If we consider the
tilt as an angle $\theta$ then we expect fractional sky variations along the slit:
$$ {\Delta S \over S}  = {1\over S} { \partial S\over \partial x } L \sin\theta $$
where $L$ is the distance along the slit in pixels and ${ \partial S/ \partial x }$
is the rate of change of the sky count $S$ with pixel $x$ in the spectrum. 
The instrument is usually critically
sampled so the PSF is 2--3 pixels. This means we expect fractional
sky fluctuations of order unity between spectrally adjacent pixels in regions
near bright sky lines. This gives:
$$ {1\over S} { \partial S\over \partial x } \simeq 1 $$

In the LDSS case the achievable rectilinear alignment is 1 pixel in 
1000 giving $\theta=0.05^{\circ}$.
In our experience this is typical of modern spectrographs as mechanical
tolerances are usually designed so that alignment is possible to $\sim$
a CCD pixel. Image distortion in the optics also turns out to be
a big effect. LDSS is a typical fast $f/2$ camera. The change in radial distortion
across a slit length will introduce an apparent rotation ($\theta'$) if the slit
is off the cardinal axes. A useful formula for this is:

$$ \theta' = {\partial D \over \partial r} {xy \over r^2} $$

for a slit at $(x,y$) wrt the optical axis axis (radius $r= \sqrt{x^2+y^2}$) where the
radial distortion $D=r_2-r_1$.

In the LDSS optics the typical distortion $\partial D/\partial r \simeq 0.02$, thus we
can estimate typical apparent rotations (using $x\sim y \sim r$)
of $\theta' \sim 2^{\circ}$. Many similar systems have fast cameras (e.g. the LRIS Keck
multislit spectrograph camera is $f/1.56$ and using the LRIS astrometry
software we find distortions of $\sim 10$ pixels over 400 pixels, 
so $\partial D/\partial r \simeq 0.02$ ) so we expect this order
of radial distortion to be typical of modern fast spectrographs.

Putting these formulas together this rotation would cause a linear sky gradient of order
30\% across a 10 arcsec slit. 
If the data could be resampled to sub-pixel accuracy to correct for tilts,
we could expect to achieve 0.1 pixel accuracy which would still
leave 10\% variations.

In principle though smooth variations can be removed. However 
another effect is slit irregularities. The milled metal slit
masks used in LDSS have 10--20 \micron\ irregularities (1 arcsec = 150 \micron\ at 
AAT's $f/8$). This is typical of machine cut masks (Szeto \etal, 1996). 
Thus we also expect $\simeq$
10\% semi-random variations along the slit due to this effect.  This can be flatfielded 
out by dividing by a dispersed white light exposure, this will be limited by flexure
between the white light and the data exposure. LDSS flexes at about 0.5 pixels$/$hour
thus we can expect a misalignment of order 0.1--0.2 pixels giving residuals of
order 1\%. 

So we are in a situation in LDSS where we are fitting slopes of order 10--30\%
with a slit length of 10--20 pixels and
with systematic variations of $\pm 1\%$. The sky lines in LDSS at low-resolution have peak
counts of $\simeq 2000$ electrons in a half hour exposure, so the random
noise will be about 2\%. Fitting along the slit would reduce this to $\ls 1\%$ at which point 
it is comparable to the systematic slit irregularities. 

How faint can we go with 1\% sky-subtraction accuracy? In
the $I$-band the sky background is dominated by the lines, if we demand an object has
$S/N\sim 3$ then the faintest that can be reliably reached, in {\em any exposure time}, is
$I_{AB}=23.6$ per arcsec$^2$. 
Fainter than that the fluctuations in the spectrum will be
dominated by sky residuals at the lines, and for low-resolution $I$-band spectroscopy
the lines occlude most of the spectrum.

How could this be improved? One crucial area with scope for improvement is the 
microroughness of the slit edges. 

\subsection{Improving multi-slit accuracy}

Conventional laser cutting (melting and vaporization) of metal (\eg Al) masks 
produces 10$-$20\micron\ roughness.  During manufacture, most metals undergo
warping during cutting which defocusses the laser.
This is one of the major sources of error in slit manufacture which in turn 
contributes to poor sky subtraction.

Recently, new slit masks made with laser-cut carbon fiber have already achieved 
an order of magnitude  improvement in edge roughness (Szeto\etal\ 1996). 
An important step by the Gemini/GMOS team (Stilburn, private communication)
was to use epoxy-bonded sheets made of 3-ply unidirectional carbon fiber with a
total thickness of only 200\micron. The center ply is orthogonal to the
outer plies, and the slits are cut at 45\deg\ to the fiber direction.
The low-power Nd:YAG laser cuts slits at 10~mm s$^{-1}$
and, remarkably, achieves a 1$-$2\micron\ edge roughness.

Let us assume an 8m 
size telescope with a larger image scale. At $f/16$ a 1 arcsec slit would be 600
\micron\ so the irregularities would be 0.1--0.2\%. The larger mirror
will accumulate more light, so we would reach this limit in a 3 hour
exposure, faster if our spectrograph was more efficient. At 0.1\% of
sky we are now observing at a surface brightness limit of $I_{AB}=26.1$ per arcsec$^2$
with forseeable multi-slit technology.
Improving the instrumental resolution will reduce the amount of spectrum
occluded by sky lines, though the peak counts in the lines will stay approximately
the same as they will stay unresolved. There will be a danger of running into
detector dark and readout noise limits.

\subsection{Nod-shuffle sky-subtraction accuracy}

It is clear that the acheivable accuracy of sky-subtraction with the nod-shuffle
technique depends on how rapidly the nod-shuffling is done. If this is done at
a fast rate changes in the night-sky background are sampled more accurately, as
well as changes in the instrument such as flexure. However characteristic timescales
for the latter are of the order of hours, so sky temporal variations will be the 
limiting factor on the residuals.

\begin{figure*} 
\psfig{figure=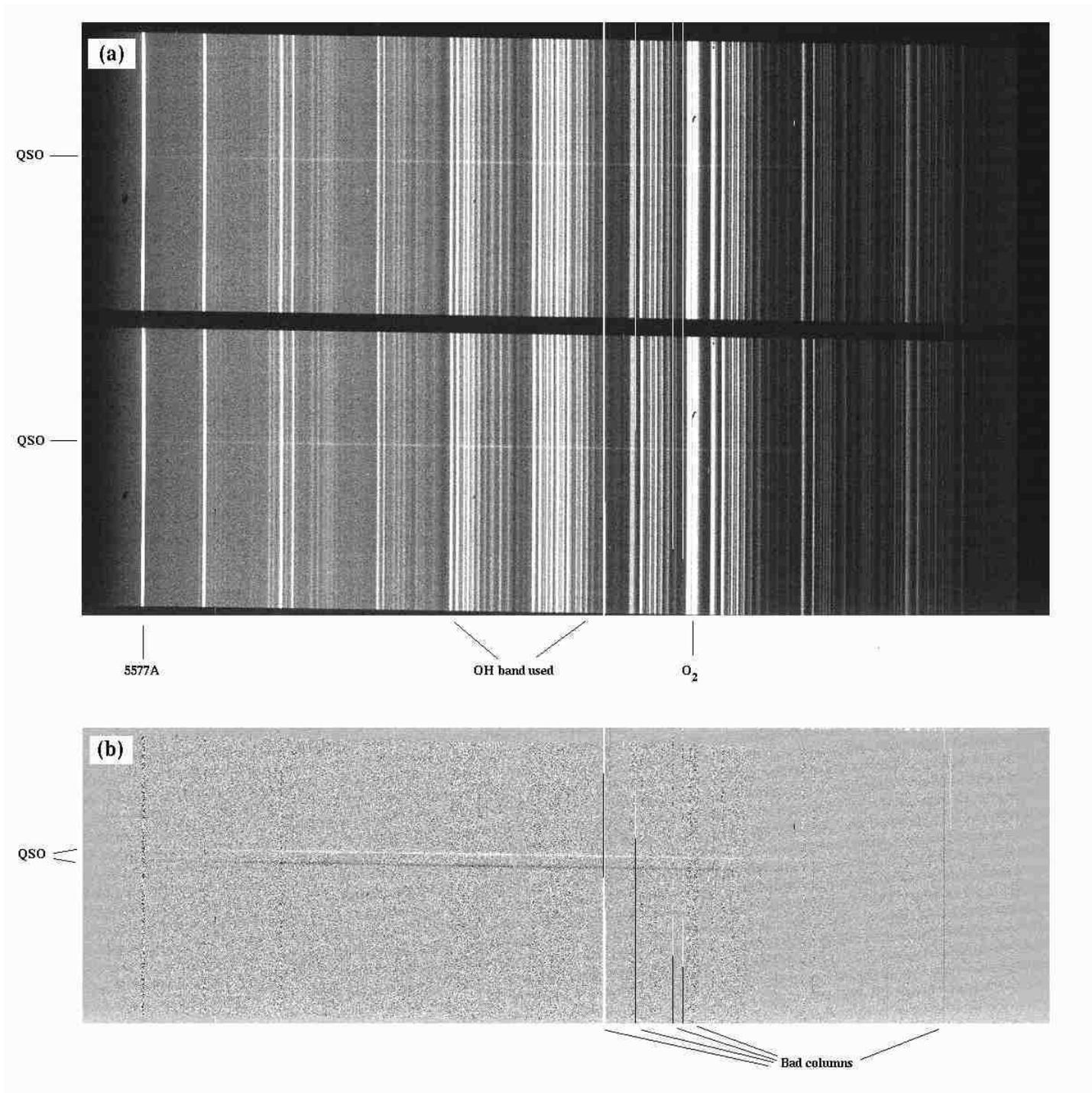,height=7.5in}
\caption{Example of the longslit data which goes into our sky-residual analysis. (a) Raw, shuffled
image (except for cosmic rays being patched out). (b) A-B subtracted image showing the 2D residuals. 
Residuals are integrated
along the slit and across a wavelength range as described in the text.}
\label{fig-examplels}\end{figure*}

In order to empirically measure the accuracy of sky-subtraction we used a sequence of
8 longslit spectra, collected on 2--3 April 2000 at the AAT in longslit mode. The
targets were faint QSOs ($I\le 22$) in a scheduled AAT science project, by arrangement
with the observers the observations were done so as to allow us to try out different
nod-shuffle times. The slit wss 4 arcmin long and the longslit
data were collected in nod-shuffle mode with the targets nodded 5--10
arcseconds along the slit. The log of the observations is given in Table~\ref{observations}.
A sample raw data frame is shown in Figure~\ref{fig-examplels}.

\begin{table*}
\caption{Log of observations for the nod-shuffle sky residual analysis, all 1800s {\em total} integration
time}\label{observations}
\def\ditto{\quad$'\,'$}
\medskip
\begin{tabular}{cccl}
\hline\hline
AAT RUN & NS-Time/secs & UT start  & Remarks \\
\hline
02APR0001   & 30    & 09:46:54  &  Some cloud (5/8$^{\rm ths}$) \\
02APR0008   & 15    & 13:06:03  &  \ditto \\
02APR0010   & 7.5   & 14:00:32  &  Clear  \\
02APR0012   & 30    & 15:18:48  &  \ditto \\
03APR0004   & 60    & 14:17:06  &  Clear, bad seeing (5--$10''$)  \\
03APR0005   & 60    & 14:50:51  &  \ditto  \\
03APR0007   & 30    & 15:40:35  &  \ditto \\
03APR0008   & 30    & 16:15:11  &  \ditto \\
05APR0007   & 300   & 14:29:03  &  Clear, v. bright O$_2$ emission (8645\AA)  \\
06APR0004   & 150   & 09:25:45  &  Clear, seeing 2--2.5$''$ \\
06APR0005   & 450   & 09:58:54  &  \ditto \\
\hline
\end{tabular}
\end{table*}

All the frames had the same total exposure time of 1800s, the only change was the rate of
nod-shuffling which we varied from as fast as 15s to as slow as 450s. Once the QSOs are
masked out the sky region of the 2D images can be used to quantify the effect of
the nod-shuffle time on sky residuals.  The data processing sequence is extremely
simple:
\begin{enumerate}
\item Frames are bias-level subtracted.
\item A median-filter smoothed version of each frame is made. The smoothing is entirely
along the spatial (Y) axis with a smoothing kernel of 21 pixels (8.2 arcsecs). Because the
slit is very closely aligned with the CCD columns ($\le 1$ pixel) and the CCD has
good flat-field characteristics this essentially replaces each pixel with a smoothed
estimate robust against cosmic-rays. 
\item The smoothed frame is used to calculate the variance map of the raw frame assuming
shot noise from the sky and the know readout noise of the detector.
\item Cosmic rays are identified as $>10\sigma$ peaks in the RAW$-$SMOOTHED map and used
to calculate an exclusion mask. Any pixel within 5 pixels of a cosmic ray peak are masked.
Cosmic ray identifications are checked visually. This mask excludes about 1\% of all pixels
on each frame.
\item The cosmic ray mask is ORed with another mask which excludes several bad columns
and the centre rows where the QSO spectra lie.
\item A sky spectrum is formed for each frame by averaging unmasked pixels along the
slit. A variance spectrum is also calculated. \label{PREV}
\item A residual sky spectrum is formed for each frame by repeating step \ref{PREV}
for the residual A$-$B frame.
\end{enumerate}
\def\dsky{\hbox{$\Delta\hbox{sky}/\hbox{sky}$}}
To calculate the fractional sky-residual \dsky\ we can integrate the residual
and sky spectra in wavelength and divide. Absolute flux calibration is not
necessary. We chose two wavelength regions: the first region encompasses the
two main OH regions in the $I$-band (7200--8880\AA) and the second region encompasses
the 5577\AA\ OI line (60\AA\ width bandpass). 
We choose to fit and remove the continuum level
from the spectrum before doing the summation. This is because there is not enough
unilluminated space on the detector to allow accurate determination and extrapolation
of the level of scattered light. In any case the integrated sky-brightness is dominated
by the lines, not the continuum, 
and it is the temporal variation of the line flux we are primarily
concerned with. Since our sky-spectrum is also integrated along 4 arcmins of slit
we can go very deep in measuring systematic residuals.

\begin{figure*} 
\psfig{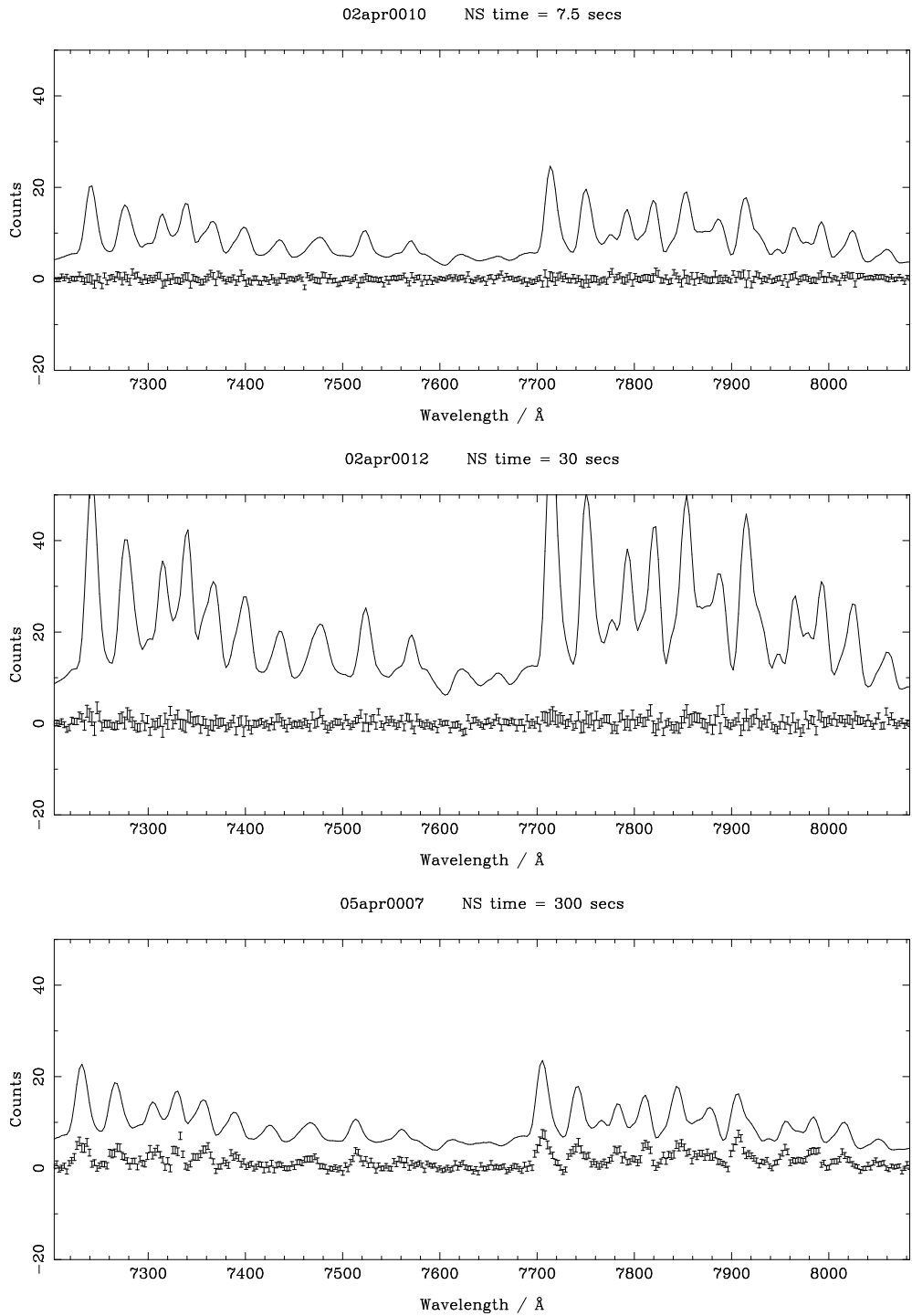}
\caption{Sky residuals in a 1800s exposure as a function of nod-shuffle time. The upper
line in each panel show the raw sky$/ 10$, the lower points (with error bars) show the
residual after nod-shuffle subtraction. A clear point-point systematic is seen in the
300s nod-shuffle exposure, while the others are consistent with zero.}
\label{fig-dsky1}\end{figure*}

\begin{figure} 
\psfig{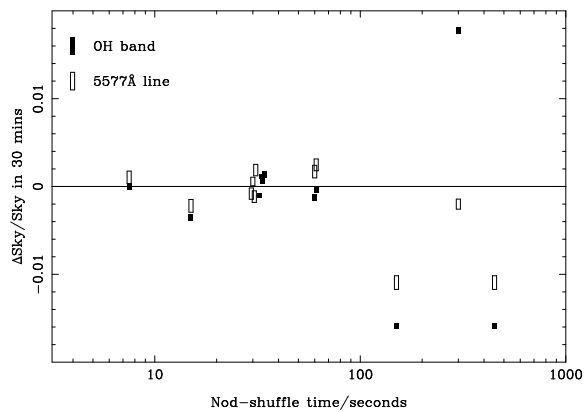}
\caption{Sky residuals vs nod-shuffle time for the chosen OH band and the 5577\AA\ line. The rectangles
indicate the range of the +/- 1$\sigma$ errors vertically and are filled for OH, open for  5577\AA. 
Small artificial offsets temporal are used at 30s and 60s to show the multiple points with clarity. }
\label{fig-dsky2}
\end{figure}

Our results are shown in several figures. Firstly Figure~\ref{fig-dsky1} shows raw and
residual spectra for our two regions for different nod-shuffle times.
Figure~\ref{fig-dsky2} shows \dsky\ plotted against nod-shuffle time for the two regions.
There is a clear trend of systematics consistent with scatter around zero at the level of
$\pm 10^{-3}$ for small nod-shuffle times ($<$100s), the level of the scatter is about
3$\sigma$. For large nod-shuffle times $>$100s there are gross systematic residuals at
the $\pm 10^{-2}$ level.

\begin{figure} 
\psfig{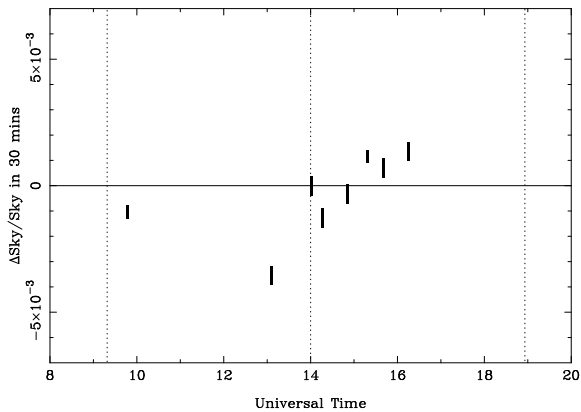}
\caption{OH sky residuals vs UT for nod-shuffle times $<$100s. 
Local midnight (14$^{\rm h}$UT) and
the end$/$start of twilight are indicated by dotted lines. We 
expect to see a negative residual
at the start of the night and the residuals should have a positive
slope with time, we do in fact see this.}
\label{fig-dsky3}
\end{figure}

One limitation of our particular nod-shuffle technique is we observe an asymmetric
seqeuence: $$ABAB ... ABAB$$ If there was a systematic change in sky-brightness during
the course of the observations we would expect to see a residual because the average $B$
frame is slightly later in time than the average $A$ frame. A systematic decrease in OH
emission  during the course of the night is often observed (Leinert \etal\ 1998). This
effect is normally explained as the result of energy stored during the day in the
respective atmospheric layers (Kondratyev 1969). We see evidence for {\em exactly} this
effect, with the correct sign, in our data (Figure~\ref{fig-dsky3}). An additional source
of long-term  variation is the effect of changing airmass  during an extended observing
sequence on a single source (Bland-Hawthorn et al.  1998). In principle it is
straight-forward to reduce these effects by improving the nod-shuffle method with a
symmetric mode, i.e.: $$ {B\over2} ABA ... ABA{B\over 2} $$ Then the $A-B$ subtraction
would cancel out any linear trend. However we have yet to try this in our AAT
implementation.

The effect of drift should also cancel to some extent for long all-night nod-shuffle exposures
which bracket local midnight.
It would be desirable to take much longer integrations with a
fast nod-shuffle rate to explore the limits of this technique. While we do not have this
data as such, what we can do is stack all our data where the nod-shuffle time is $<$100s.
This gives us a 5.5 hour very deep exposure, albeit with a variable nod-shuffle time. The
residual point from the 5.5 hour stack is 
$(4.0\pm 1.2)\times 10^{-4}$ --- a $3\sigma$ detection. It is important to realise that this
is an impressively small residual corresponding to a $I_{AB}=28.3$ mags arcsec$^{-2}$ source. This level of accuracy
is a factor of 10--20$\times$ better than is typically achieved with slits (see
Section~\ref{sec-sky-mslit}).

We also emphasize that this is a lower limit to what could be achieved with faster
nod-shuffle times. One could nod-shuffle faster (e.g. 10s) for a whole very long
exposure. Also one should implment the symmetric mode to cancel long-term
sky-brightness drifts. Finally for the ultimate
sky-subtraction limits one could combine nod-shuffle {\em with} slits to allow 
for 2D interpolation and removal of any local residuals after nod-shuffle
subtraction. Accuracies of $10^{-4}$ or better should be achievable.

\subsection{Comparison of residuals to theorectical predictions} \label{sec-theory}

We have shown that the nod-shuffle residuals appear to 
be characteristically smaller for nod steps below 100~sec 
compared to longer 
sample exposures. We now examine this with a simulation of
the nod-shuffle technique using a model which attempts to
describe the time-variable behaviour of OH emission. 

Suitable observations for deriving the temporal power spectrum 
of OH are hard to come by. Line strength variations on timescales
of 5--10 mins are given by Bland-Hawthorn\etal\ (1998) for optical
lines and Ramsay\etal\ (1992) for near-infrared lines. The latter
reference shows the OH behavior to be approximately sinusoidal on timescales
of an hour with a peak-to-peak amplitude of about 10\%. On longer
timescales, the OH variation is more erratic.

Our model for atmospheric variability uses a finite set of sinusoidal modes
with periods 16, 23, 26, 29, 38, 51 and 101 mins. The amplitude of the 
variations are inversely related to the period such that the 16 min dominates,
in rough accordance with the wave-like structures observed by
Ramsay\etal\ (1992). The peak-to-peak amplitude is
15\% of the mean line strength. For each mode, there is a 5\% dispersion in 
the period and amplitude, each with random phases. Our predicted behavior is 
in good agreement with the above references. 

\begin{figure}
\psfig{figure=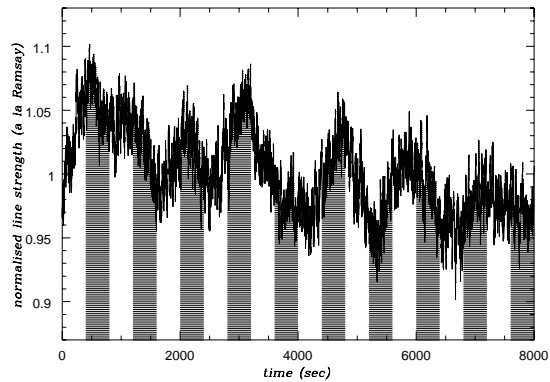,width=\hsize}
\caption{Example OH airglow time series generated from our model. The shaded bands
indicate periods of 400 secs.}
\label{fig-oh-series}
\end{figure}

However, high cadence observations show clear evidence for stochastic behavior 
on shorter observational timescales. Here, we found data from the 2MASS Wide-field Airglow 
Experiment\footnote{See: http://pegasus.phast.umass.edu/2mass/teaminfo/\
airglow.html}
to be the most useful (Adams \& Skrutskie 1997). The H band observations
have an order of magnitude finer sampling than in Ramsay\etal\ (1992).
We simulate this by including a component of $1/f$ noise within our model
(cf. Barnes \& Allan 1966). To generate the $1/f$ component we use
gaussian white noise scaled to 5\% ($1\sigma$) of the mean line strength 
(see Adams \& Skrutskie 1997, Fig.~2)  convolved 
with Green's impulse function ${\cal I}(t-t_o) = c(t-t_o)^{-0.5}$ ($t > t_o$); 
${\cal I}(t-t_o) = 0$ ($t \leq t_o$) . For convenience, we set $c=1$ and
sample the time axis in units of seconds. An example time series is shown
in Figure~\ref{fig-oh-series}.

\begin{figure} 
\psfig{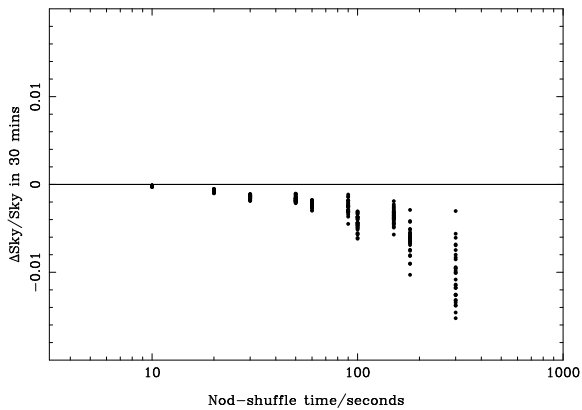}
\caption{Simulation of the expected residuals in a 1800s exposure for
different nod-shuffle times. This should be compared with Figure~\ref{fig-dsky2}.}
\label{fig-oh-sim}
\end{figure}

In Figure~\ref{fig-oh-sim}, we have attempted to simulate nod-shuffle sampling of our
model atmosphere. The total exposure time is 1800 secs and the time series is
sampled at all possible time steps (longer than or equal to 10~sec) that
lead to an integer number of cycles. For each nod exposure, the simulation
was run 10 times. The mean residuals (and $1\sigma$ errors) are shown as
a function of the nod exposure. There is a evidence for a change in character 
on either side of about 2~min time steps. The residuals with 2~min samples
or longer are $10^{-3}$ or larger; the residuals from faster sampling are 
$1-4\times 10^{-4}$.

Repeated runs of our model atmosphere show that this changeover can be as 
short as 1~min. There are also times when short sample time steps lead
to big residuals (e.g. 20~sec) and when long time steps lead to residuals
smaller than $10^{-3}$. These are times when the nod-shuffle sequence happens
to fall in or out of step with a beating atmosphere. Airglow is
clearly a complicated phenomenon: empirically it is clear that
the nod-shuffle time should be $\ls 30$ sec. The total number of
shuffles should not greatly exceed $\approx 10^2$ per readout if one is to avoid 
significant degradation from trapping sites within the silicon substrate
(Bland-Hawthorn \& Barton 1995). Given the periodic nature of the airglow
oscillations it is possible that an optimal shuffle sequence ought to 
have variable time sampling to avoid beating.

\subsection{Object-sky balance}

The question arises what is the optimum balance between OBJECT and SKY time in a 
nod-shuffle sequence? This especially important when we are nodding out of a microslit
and the SKY frame is  not collecting any object photons. Perhaps one should cut down on
the relative frequency of SKY frames? It turns out the optimum balance is in fact 50:50,
i.e. symmetrical. Consider an exposure of total time $T$ where a fraction $x$ is spent on
OBJECT and $(1-x)$ on SKY. Let $O$ and $S$ be the object and sky flux
(photons$/$pixel$/$sec). We will neglect readout noise which is equivalent to assuming
that $T$ is long enough that both $O$ and $S$ are large enough that their shot noise
dominates over the readout noise, which is optimum. We will also assume that the object
is much fainter than the sky, i.e. $O<<S$.

We form the residual sky-subtracted image as:
$$OBJECT - {x\over(1-x)}\, SKY $$

Then the signal to noise in the residual image is:
\let\ds=\displaystyle
\def\half{1/2}
\begin{eqnarray*}
S/N &=& { xOT \atop \sqrt{STx \left(1+{\ds x\over\ds(1-x)}\right)} } \\
    &=& {OT^{\half}\over S^{\half}} \sqrt{x(1-x)} \\
\end{eqnarray*}

$x(1-x)$ has a maximum when $x=0.5$, i.e. equal times on OBJECT and SKY. $x$ could be
reduced in a scheme where the SKY frames were averaged over mulitple observations or
multiple slits before subtracting, however one then loses the crucial ability of the
simple nod-shuffle scheme to follow precisely short-term and long-term temporal
variations in the sky and eliminate local effects such as flat-fielding, fringeing,
flexure, slit roughness, etc., from the sky subtraction.  

Finally we note that, not surprisingly, 
in the case $O>>S$, i.e. the object is much brighter
than the sky, the maximum $S/N$ is obtained when 
as much time as possible on the OBJECT. However 
in this regime the sky contribution to the statistical noise is negligible
so nod-shuffle is not very useful, except possible in a observation where 
systematic effects were an important concern (for example velocity dispersion
measurements of bright galaxies as discussed in Sembach \& Tonry, 1996).

\subsection{Effect of random objects on sky-subraction}

We conclude our section on sky-subtraction by considering the
effect of random interloping objects on the accuracy. In our
simple AAT implementation we nod between two positions, so there
is some chance there will be an interloper in the sky position.

We can estimate this effect using deep galaxy number-magnitude counts
(Hogg \etal\ 1997). At our HDF-S limit of 
$R=24$ there are $\sim$ 60000 galaxies deg$^{-2}$
which equates to a 1 in 200 chance of a $\sim $ 1 arcsec$^{-2}$ aperture encountering
one. This is consistent with our HDF-S observations where two negative spectra
were observed. 

This can be alleviated by dithering the sky position. This can be done
in two ways. Firstly seperate nod-shuffle exposures can have {\em different}
sky positions. Then the frames can be combined with outlier clipping
{\em after\/} pair-subtraction to effectively remove the interloping spectrum
with negligible effect on signal$/$noise (as only a tiny fraction of
pixels are rejected).

Secondly a more technically
sophisticated approach would be to drive to a {\em different}
sky position on each shuffle. This would be advantageous for short
shuffle runs where there are not many individual exposures. A disadvantage
is that the effective average sky is not outlier clipped, however the
flux of interlopers is still greatly reduced. 
We note this mode is not possible with our AAT system,
but is in principle straight-forward to implement.

In view of the remarks in Section~\ref{sec-ultradeep} about 30m telescopes
it is useful to consider the ultimate achievable limits. For very faint galaxies
it would be sensible to use smaller slits, because the
faintest observed objects in the Hubble
Deep Field typically have half-light radii of only 0.1--0.2 arcsecs
(Gardner and Satyapal, 2000). At this limit ($I_{AB}\sim 30$) there
are of order $\sim 10^6$ galaxies deg$^{-2}$, so the covering
factor at 3 half-light radii is still only 10\%. Thus the
sky-subtraction problem is still tractable with dithering.

Finally we note even with an interloper the sky-subtraction itself is
still accurate. This contrasts with the longslit case where the
interloper can disturb the interpolation. The result is the sum 
of the positive and negative spectrum, if the relative brightnesses
are similar and the signal$/$noise is sufficient in principle
redshifts can be derived for {\it both} objects.

\section{Sample observing modes} \label{sec-mode}

A discussion of the different modes of observing whch have been tried
with LDSS++ is useful to show the potential new capabilities.

The most conventional mode is multi-object spectroscopy with wide wavelength
range. Sample raw data was shown in Section~\ref{implementation}.

We would like to illustrate briefly two other modes which have
been  used recently to achieve very high multiplex levels of 1000--2000
objects per LDSS mode.

\begin{figure*} 
\psfig{figure=fig10a.ps,width=5in,angle=-90}
\hbox{
\raisebox{1.6cm}{\psfig{figure=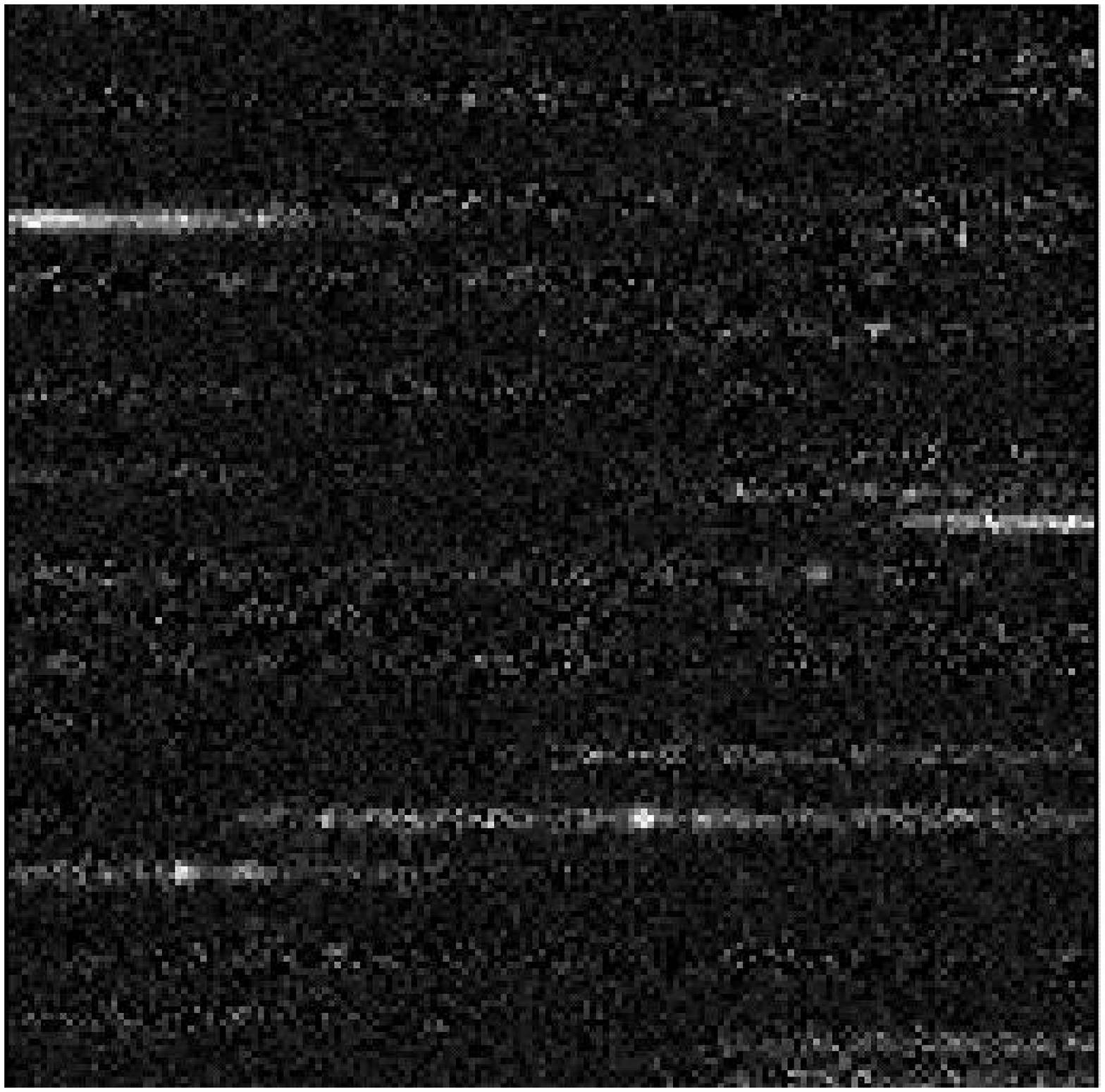,angle=-90,width=3in}}
$\stackrel{\psfig{figure=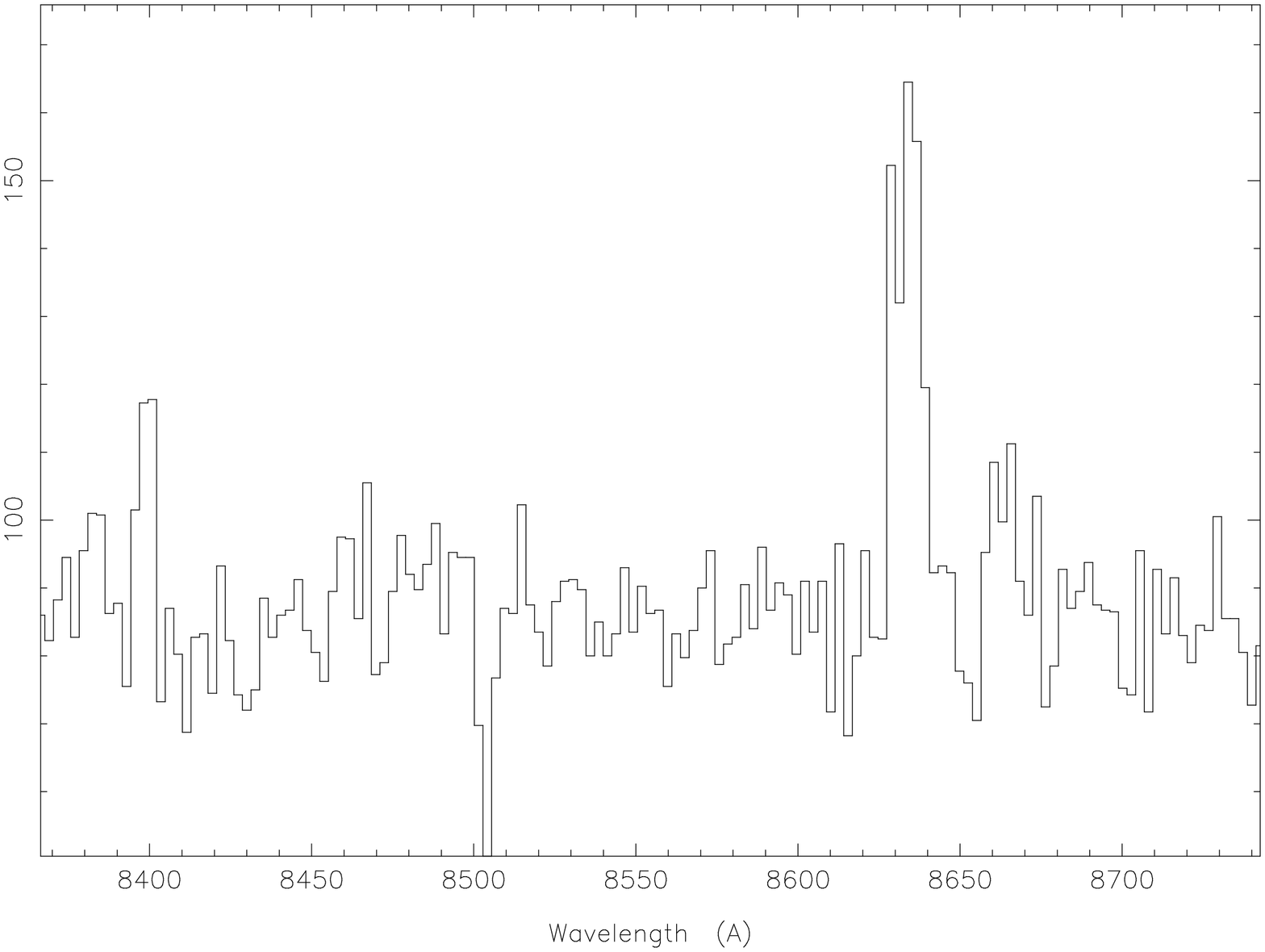,width=2.5in,angle=-90}\\ \\ \\}
{\psfig{figure=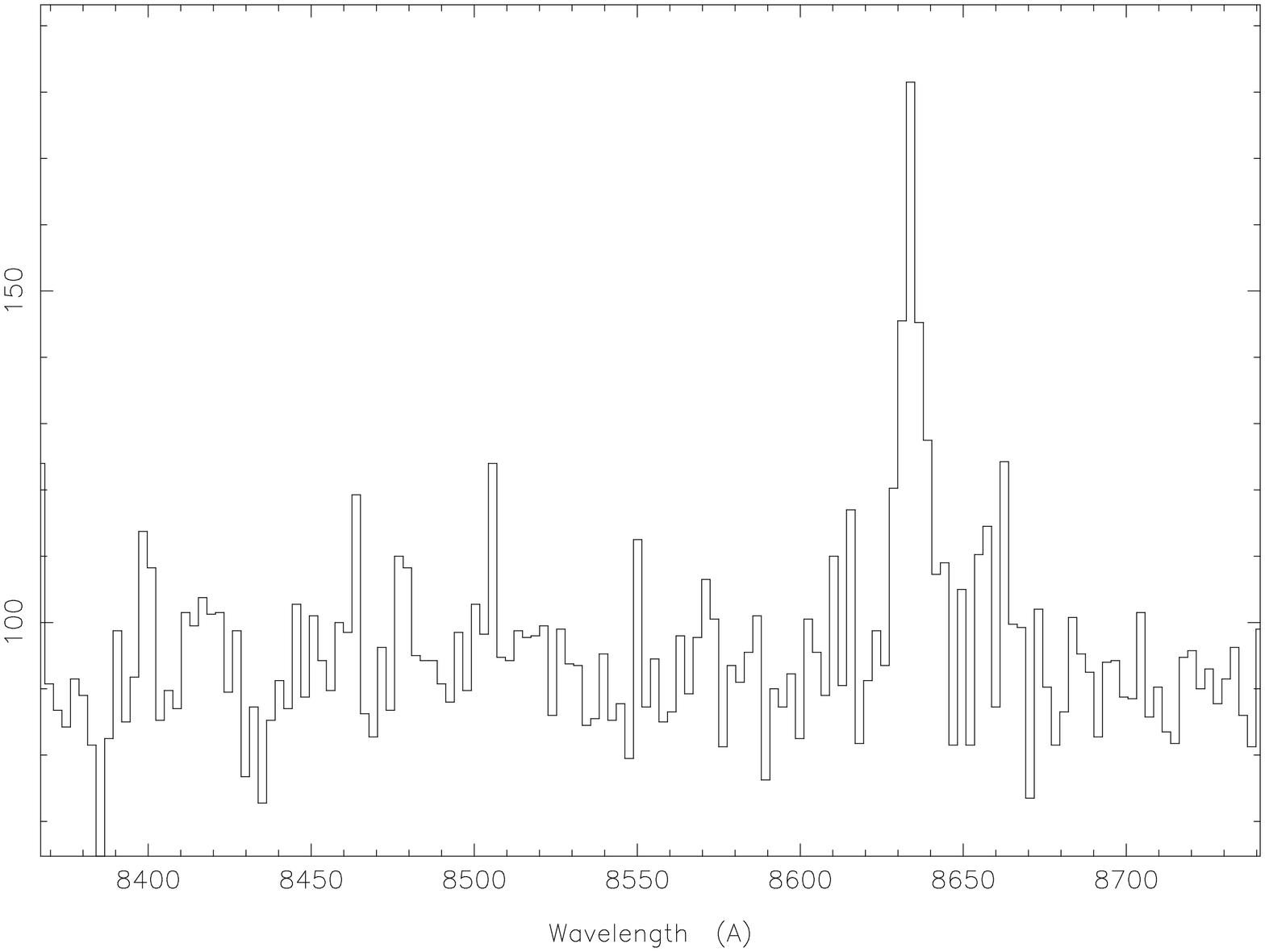,width=2.5in,angle=-90}}$
}
\caption{\Halpha\ spectroscopy of the $z=0.32$ cluster
AC114. Top: layout of the spectra in the 9 arcmin
FOV, Bottom left: Zoom showing sky-subracted, dispersed image
with a couple of \Halpha\ lines visible. Bottom right:
Two sample extracted spectra showing \Halpha\ and [NII]. 
}\label{fig-ac114}
\end{figure*}

It is well known that use of a blocking filter to limit the
wavelength range of a spectra allows many more slits to be
used on a mask without spectral overlap. When
this technique combined with the use of microslits an extremely
large multiplex results and allows high-density mapping of
fields in chosen spectral lines. For example in the last year
LDSS++ has been used to map \Halpha\ emission in the core
and outskirts of the $z=0.32$
galaxy cluster AC114 (Couch \etal\ 2000). The TAURUS blocking filter R6
was used which gives a bandpass of 400\AA\ for \Halpha\
(and [NII]) at the cluster redshift. Using this technique 828 slits
were placed on galaxies in a 8 arcmin field around the cluster.
Figure~\ref{fig-ac114} shows a diagram of the spectral layout
on the detector, it can be seen that despite
the large number of slits and good 2D coveragre of
the cluster no overlap occurs. Also shown is
a zoom are actual sky-subtracted cluster spectra where the
\Halpha\ lines can be seen. 

Another mode which has been developed for LDSS++ takes the multiplex
to an extreme limit by taking advantage of the superb sky-subtraction
without a slit. The key idea is to place microslit apertures on large
numbers of targets (up to several thousand) without regard to spectral
overlap, and possibly even without a blocking filter.

Of course the dispersed sky from such a configuration will
generate a very complex, overlapping pattern.  However this
can still be removed by the nod-shuffle technique, and the
residual noise level can be easily calculated. Any features
left can have a measurable significance assigned to them.

\begin{figure*} 
\psfig{figure=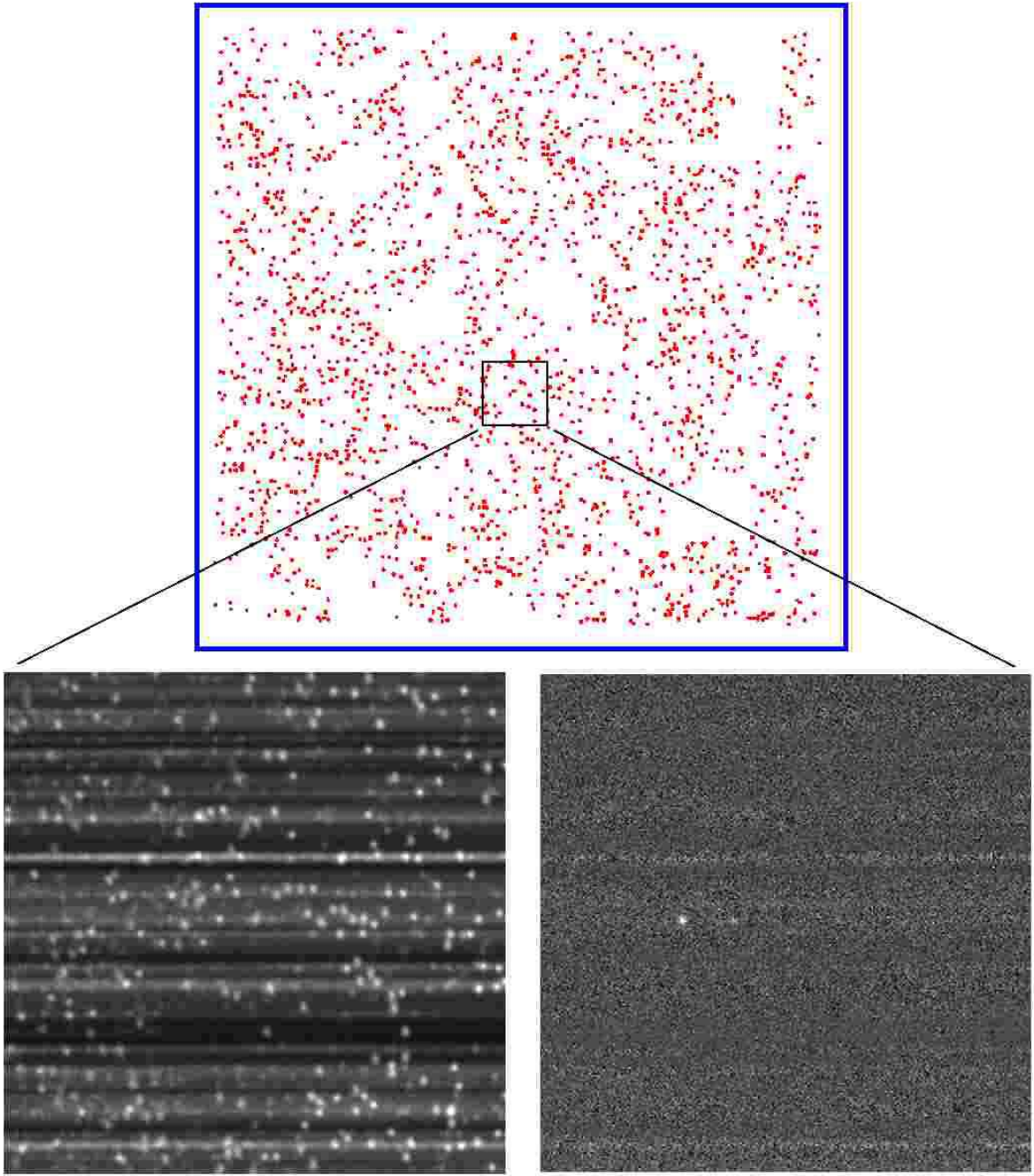,width=\hsize}
\caption{Illustration of a region of data 
in the `Pseudo-slitless' mode. The full mask (about 7 arcmin)
is shown at the
top. Slits have been placed on every object with $R<26$ (except
near bright foreground stars). The slit density
is about 50 arcmin$^2$. 
The lower panels show a region about 1.6 arcmin across zoomed in.
Left: before sky-subtraction showing the complex overlapping pattern.
Right: after sky-subtraction showing a noise pattern plus some bright
emission lines from a low-redshift galaxy. Continuum from some bright objects
can also be seen.}\label{fig-ps}
\end{figure*}

Why would such observing be useful? Well one example project
is illustrated in Figure~\ref{fig-ps}. Here $\sim 2000$ slits
were placed on galaxies selected to $R\sim 26$ 
in a 7 arcmin field called the `Herschel Deep Field' 
(McCracken \etal\ 2000). The sky is removed by nod-shuffle
and a noise map is calculated. If a galaxy has strong emission
lines then they peak up above the noise map. 

Essentially we are searching virtually all galaxies in the field
for emission --- so it is similar to a slitless grism survey. However
we still have a mask in the beam so the level of the sky background
is enormously reduced (a factor of 50 in this case) with corresponding
increase in signal:noise. Because of the similarity we call this
method `pseudo-slitless'. Another way
of looking at this is we are using our prior knowledge of where galaxies
are in the broad-band image to exclude unwanted sky photons. The background
is higher  than conventional spectroscopy, but more
objects are observed simultaneously. In principle these effects cancel
exactly, if there are $N$ times more microslits then the average background
is $N$ times higher and the exposure has to be $N$ times longer for the
same signal:noise. In practice there are gains in efficiency due to
factors such as overlap and clustering which complicate slit assigment
in the normal case. For the real example in Figure ~\ref{fig-ps} the
factor $N\sim 10$.  

How does this approach compare against, for example, narrow-band imaging
and scanning in wavelength? In the pseudo-slitless mode we are pre-selecting
from the broad-band so it is possible to miss pure-emission line objects. 
If we ignore this difficult to quantify
handicap then there is a net gain. Let us assume
the tunable-filter instrument has the same absolute throughput as
the spectrograph. The pseudo-slitless approach gives a very large wavelength 
coverage --- in our example 5300\AA. At a resolution of 20\AA\ then
that is needs 265 tunable filter settings. In our example the pseudo-slitless
approach has 10 times higher background --- so the gain is a factor of
$\sim 26$, {\em for the objects searched}. 

Some data was collected in this mode in August~1999. The project is
attempting to quantify the space density of \Halpha, \Hbeta, [OII], Ly$\alpha$ 
line sources at $z=0.2$, 0.6, 1.1, 5.6 respectively (Glazebrook \etal\ 2000b).
 
There is of course an inherent ambiguity: if an emission line is
detected how can we determine which microslit it came from? There
will be many candidates along it's dispersion track. This is
resolved in two ways: firstly a minimum separation is enforced
between slits (e.g. a few arcsec) to allow for errors in the
traceback. Secondly the observations are made for different mask
orientations on the sky. As the grism is kept fixed we get a different set of
tracks. For the observations here positions of 0$\deg$ and 180$\deg$
were used: the emission line is dispersed in opposite directions
in each case and the correct microslit lies halfway between them.

Finally we note that it is possible to arbitrarily combine the approaches 
described here. For example in the pseudo-slitless mode blocking filters
can also be used: this will limit the spectral coverage but also reduce
the background. There is a choice as to whether to go for low or high
microslit densities --- the latter will mean having to deal with
confusion and a higher background.

\section{Future prospects} \label{sec-future}

\subsection{Nodding with infrared arrays}

\subsubsection{Prospects for mimicking shuffling directly}

Can the nod-shuffle concept be extended to include IR-sensitive devices?
We have been asked this question many times ---
since the OH night sky lines account for 98\% of the sky background in the 
J and H bands this would give major gains. 
However, infrared arrays are fundamentally different devices from CCDs.  
In conventional arrays, the pixels are not charge-coupled 
so that charge cannot be shifted between pixels (Rieke 1994, McLean 1997).

CCDs are monolayer devices where the charge is normally
shifted row by row into the read-out (shift) register. Pixels within
the read-out register are read out serially towards the output
amplifier by means of 2, 3 or 4-phase shift electrodes.
In contrast, the Rockwell hybrid arrays are 2-layer devices which use 
a thin HgCdTe film to collect the light, which in turn is connected 
pixel-by-pixel via indium bump bonds to a MOSFET multiplexor. Each
pixel is addressed in an $(x,y)$ fashion through the use of a row
and a column shift register at two edges of the multiplexor.
In the `source-follower' multiplexor design, the bump bond makes contact 
with a MOSFET.  When IR photons hit the light-sensitive layer, the 
electrons are transferred through the bond to the capacitance-storing 
MOSFET gate. This gate is bordered by a `source' (grounded) and `drain'. 
This circuitry allows for a `non-destructive read' (NDR) of the voltage 
across the gate.  Another FET is attached to the gate to allow every 
element of the array to be `reset' in a single action.

We have considered possible modifications to the IR array design which  would 
allow for the equivalent of CCD-style charge shuffle operations, i.e. that 
contains  two or more switchable pockets per pixel in which to store charge.  
Unlike Rockwell arrays, there exist multiplexors which use arrays of FETs as 
op-amps which simply transfer photogenerated charge to an integrating capacitor
(e.g. Kozlowski 1996). 
One could conceive switching between a pair, or more, of  integrating capacitors 
in which to build up charge sequentially over time.

However, the more connections you attach to the detecting  node, the more the 
capacitance goes up, and therefore the read noise.The array multiplexor already 
has a higher circuit density compared to CCDs and this would increase it further. 
This would be a very difficult technology to develop.

\subsubsection{Can one use Non-Destructive Reads to facilitate beamswitching?}

\begin{figure}
\psfig{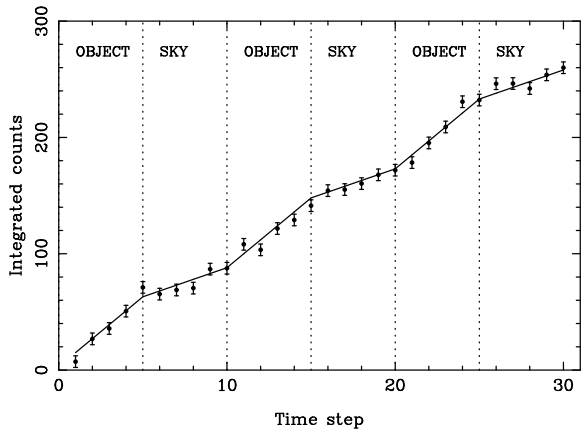}
\caption{Illustration of the `IR nod-NDR concept'. As counts are accumulated in the
NDR mode the telescope is switched between OBJECT and SKY periodically. A double-slope
least-squares fit is performed to derive the OBJECT and SKY count rates. It turns out
that this is not useful (see text).}
\label{fig-ir-nod-ndr}
\end{figure}

We have also considered the question of
whether the non-destructive read mode with
ramp sampling could be used to mimic shuffling, for
example by switching between OBJECT and SKY while sampling up
the slope and solving for OBJECT and SKY
count rates simultaneously while still allowing readnoise reduction (the
main point of ramp sampling). This is illustrated in Figure~\ref{fig-ir-nod-ndr}.

We have solved analytically the case for double-slope least square fitting.
If $n$ is the total number of reads with error $\sigma$ 
we find for {\em large }\footnote{Full derivation is available on request
from the authors} $n$ that the error on
the OBJECT slope $\sigma_o$ is given by:
$$\sigma^2_o = {48 k^2 \sigma^2 \over n^3 \Delta t^2 }$$
where $k$ is the number of OBJECT-SKY sub-intervals (e.g. $k=6$ in 
Figure~\ref{fig-ir-nod-ndr}). If we compare this with the classic
single least square formula ($\sigma_o = \sqrt{12}\sigma^2 / n^3 \Delta t^2$), 
we derive the ratio:
$${\sigma_o (\hbox{double)} \over \sigma_o(\hbox{single)}} = 2k $$
The factor of 2 is the usual beamswitching factor encountered in 
Section~\ref{sec-mplex}. We see the effect of beamswitching is to 
increase the noise in proportion to the number of switches, this is
because the switching reduces the baseline for slope fitting. It turns
out for reasonable values of $n$ and $k$ this is not a useful technique.
For example suppose the array can be read out every second during a
1800 sec exposure. Single least-squares would give a noise reduction of
$\sim 12\times$, if we then beamswitch every 30s this becomes a noise
{\em increase} of $\sim 4.9\times$.

Finally we note from Section~\ref{sec-theory} that in any case the
assumption that the source is of constant brightness and that counts
$\propto$ time is very dubious for the sky. The airglow is a stochastic
phenomenon with a lot of variation and will deviate from a linear
growth. This will generate artificial noise in a line-fitting approach,
even with the classical single-line fit. NDR slope-fitting has become
a standard technique at many observatories, but the effects of sky-background
variations on noise have not been studied.

\subsubsection{Physical array shifting}

The most reasonable option for mimicking something like charge shuffling is
to form two adjacent images at the detector either by nodding the collimator or
by a physical movement of the array. 
The present IR arrays are 1024$\times$1024 pixels in size, although Rockwell
are expected to produce 2048$\times$2048 formats in the near future.
`Detector nodding' is much the preferred option for a number of reasons. 
First, a nodding collimator leads to different light paths for the object 
and sky positions. Secondly, in infrared instrumentation, the collimator
must image the pupil onto the cold stop with care. Thirdly, the physical
tolerances at the collimator are made much tighter by any focal reduction 
compared to the tolerances of detector movement.  Finally, the array has 
much the lowest mass of any component of the system, and a 1~Hz movement 
through a few millimeters is not an excessive strain on the electrical 
bonds.

An advantage of IR `shuffling' over optical shuffling is that
the stored charge is not subject to trapping sites. Furthermore,
the detector needs only to be partitioned into two panels 
rather than the three panels of optical CCDs. A distinct disadvantage is
that in IR shuffling the flatfield structure will be different in
OBJECT and SKY regions. However this effect can be averaged out by swapping
the OBJECT and SKY positions on the array between successive exposures.

For a detector with 18$\mu$m pixels, the physical movement of the array 
should be accurate to better than 2\% of a resolution element (assumed 
to be 3 pixels).  Precision movement to this level is routinely achieved
in, say, a mechanism for optical focussing. But within a cryogenic 
environment, 1$\mu$m accuracy presents a moderate challenge. This seems 
feasible with 
either a linear variable differential transformer (LVDT) or a linear 
encoder.  Piezo-electric control at cryogenic temperatures is a more 
difficult prospect.  We note that a well sampled resolution element 
(say 5 pixels width) may in fact allow wavelength calibration to 
sufficiently high accuracy between the object and sky exposures that 
the precision can be relaxed by post-analysis. However, data analysis 
is greatly simplified by the ability to remove sky accurately by 
straight subtraction since no interpolation is required.

\subsection{Applications to non-contiguous spectroscopy}

The nod-shuffle technique allows accurate sky-subtraction without
requiring sky spectra which are spatially contiguous on the detector and the sky.
Thus it is particularly suitable for non-contiguous optical systems
such as fiber spectrographs and integral field unit spectrographs (IFUs), both
fiber based and non-fiber based.

Application of fibers to faint spectroscopy
have been limited by sky-subtraction accuracies
of typically 3\% (Wyse \& Gilmore 1992), which are due
to variable fiber transmission. The nod-shuffle technique
can be applied to fiber spectrographs providing there is spare
room on the detector as outlined earlier, the 2D shuffled subframe of SKY
spectra through the fibers is simply subtracted from the 2D OBJECT subframe.

Due to the quasi-simultaneity the effect of varying fiber throughput,
which varies on a much long timescale (hours),
will cancel out as the sky is observed through exactly the same fibers.
At the AAT we have already experimented with nod-shuffle
using the Two Degree Field fiber spectrograph and have obtained
shot noise limited subtraction implying systematics $<<1\%$ (Glazebrook \etal\
1999).

The application to IFU's is also straight-forward. Accurate sky-removal
is achieved by subtracting the shuffled frames before individual IFU
element spectra are extracted and assembled to make a data cube. Just
like the slit case the object could be nodded between two positions on
the IFU, or the nod throw could be large enough to move the whole IFU
to clear sky. While the effect of calibration of elements on
sky-subtraction is eliminated, it must still be solved if accurate
spectro-photometry is desired.

\subsection{Ultra-Deep spectroscopic exposures}\label{sec-ultradeep}

The promise of nod-shuffling is of course that the extreme precision
of sky cancellation will allow very very long deep spectroscopic exposures.
It is interesting to compare ground-based spectroscopy with space astronomy
(X-ray, IR, etc.). In the latter it is common to see total exposures of many days
to weeks in total, whereas in the former it is rare to see total exposures
of longer than a night's observing. 

Why is this? The answer is because of
the high sky background then one reaches a limit in only a few hours observing
where one is dominated by the systematics of how well one can remove it. This is
doubly important because the sky spectrum exhibits extraordinarily complex
structure. Also as we have seen in Section~\ref{sec-sky-mslit}  
there are a large number of {\em seperate} instrumental effects 
which {\em all} operate at the $0.5-1\%$ level.

The beauty of the nod-shuffle technique is that it is a perfectly differential
experiment and all of these effects are removed simultaneously from the
sky-subtraction process. They still affect the object spectrum, but that is
far less important compared to the random noise.

The question arises then will the nod-shuffle technique permit the use of
ultra-deep exposures, lasting $10^5-10^6$ secs, for optical spectroscopy?
We believe it can. At the level of sky-subtraction precision demonstrated we estimate that
one could obtain a good spectrum of a $I_{AB}=27.2$ galaxy 
(i.e. $3\sigma$ above the sky limit).
At a resolution of $R\sim 800$ one could reach this in a 200,000 sec exposure (7 nights)
on a 10m telescope with a 35\% efficient spectrograph. Using microslits one could
squeeze many parallel targets into even a small field.

We re-emphasize that we believe
our current sky-subtraction accuracy is only an upper limit to what can be achieved.
The sky-background can be reduced further by observing at higher-resolution so
the OH lines do not dominate the spectrum. The intra-OH continuum variations may be
far less rapid so even
greater accuracy could be obtained.\footnote{On the basis
of laboratory experiments (Abrams \etal\ 1994), there may exist fainter 
rotational-vibrational bandheads in between the bright OH bands, in which 
case the intra-OH `continuum' varies on the same timescales as the rest 
of the OH emission. However, the actual contribution of these putative
features to the intra-OH background light remains highly uncertain.} 
Assuming we could reach $10^{-4}$ of the
sky at a resolution $R=5000$, then faintest object would be $I_{AB}=29$. This could
be reached in a $10^8$ second exposure (3 years!) or a more reasonable $10^7$ exposure
if the spectrum was post-hoc rebinned to $R=500$. 

30--50m telescopes are being planned at the time of writing, these would reach the same
limits an order of magnitude faster. We emphasize that {\em without nod-shuffle}, or
equivalent, techniques these telescopes would reach the systematic limit for
spectroscopy in a mere one hour exposure!

\section{Summary and Conclusions}

We have explained the virtues of the nod-shuffle technique for CCD-based optical
spectroscopy: we reach a new level of sky-subtraction precision of 0.04\%.
This is in accord which predictions from a reasonable physical model of 
atmospheric airglow.

This technique also permits a great increase in the multiplex gain of multi-slit
spectrographs we have quantified those gains and showed that they are the most
in high-object density regimes.

We have outlined our thoughts on IR techniques equivalent to nod-shuffle. Possibly
new circuit designs would allow charge storage but would need to be developed. Given
the importance of IR spectroscopy on future large telescopes the scientific case
for doing so is strong. Failing this we have outlined a less satisfactory, but
still useful concept, for physically moving the IR array.

For very large telescopes (10m and greater) the precision of sky-subtraction is
a real barrier for ultra-deep spectroscopic exposures. The systematic limit of
ordinary slit subtraction is reached in only a few hours. The nod-shuffle technique
offers a remedy, and promises the possibility of extremely long
exposures, it's ultimate performance remains to be explored.

\acknowledgements{ 
{\bf Acknowledgements: } We would like to thank Warrick Couch, Richard
Bower and Collaborators for permission to show AC114 data and Pseudo-slitless data. We
thank J.R. Barton, L.G. Waller \& T.J. Farrell for their hard work on the 
implementation of nod \& shuffle
at the AAT and the AAO Director, Brian Boyle, for supporting this research with personnel
and financial resources. We acknowledge helpful conversations with D. Hall and J.
Stilburn and the helpful remarks of the anonymous
referee. Thanks also go to Gavin Dalton and Julia Kennefick for allowing 
us to use their AAT time to facilitate the nod-shuffle tests described in this paper. 
The night-sky is acknowledged as a worthy opponent.}

\newpage


\begin{references}

\reference{} Abrams, M.C., Davis, S.P., Rao, M.L.P., Engleman, R., Brault, J.W. 1994,
ApJS, 93, 351

\reference{} Adams, J.D. \& Skrutskie, M.F. 1997, preprint

\reference{} Barden, S.C., Arns, J.A. \& Colburn, W.S. 1998, In Optical Astron.
Instrumentation, Proc. SPIE, 3355, 866

\reference{} Barnes, J.A. \& Allan, D.W. 1966, Proc. I.E.E.E. 54, 176

\reference{} Bland-Hawthorn, J., 1994, Anglo-Australian Tunable Filter, internal 
document (see http://www.\
aao.gov.au/local/www/jbh/ttf/docs/aatf.ps.gz)

\reference{} Bland-Hawthorn, J., 1995, In Tridimensional Spectrscopic Methods in
Astrophysics, ed. G. Comte\& M. Marcelin, ASP Conf., 71, 369

\reference{} Bland-Hawthorn, J. \& Barton, J. 1995, AAO Newsletter 74, 10

\reference{} Bland-Hawthorn J., Veilleux S., Cecil G. N., Putman M. E., 
Gibson B. K., Maloney P. R., 1998, MNRAS, 299, 611

\reference{} Bland-Hawthorn, J., Glazebrook, K., Barton, J.R., Waller, L.G. \& 
Farrell, T.J. 2000, in preparation

\reference{} Blouke, M.M., Yang, F.H., Heidtmann, D.L. \& Janesick, J.L. 1988, In
Instr. for Ground Lased Optical Astron., Present and Future, Proc. SPIE.

\reference{} Comte, G., Marcelin, M., 1995, Tridimensional optical spectroscopic 
methods in astrophysics, Proceedings of the IAU Colloquium 149,
ASP Conference series Vol. 71.

\reference{} Couch, W.J., Balogh, M.L., Bower, R.G., Smail, I.S., Glazebrook, K.,
Taylor, M., 2000, ApJ, in press


\reference{} Cuillandre J. C., Fort B., Picat J. P., Soucail J. P., Altieri B., 
Beigbeder F., Duplin J. P., Pourthie T., Ratier G., 1994, A\&A, 281, 603

\reference{} Dressler, A. 1984, ApJ, 286, 97

\reference{} Fossum, E.R. 1997, IEEE Trans. Elec. Dev., 44, 1689

\reference{} Gardner J.P., Satyapal S., ApJ, in press

\reference{} Glazebrook, K., Ellis, R. S., Colless, M. M, Broadhurst, T. J., 
Allington-Smith, J. R., Tanvir, N. R., 1995, MNRAS, 273, 157

\reference{} Glazebrook K. 1998, AAO Newsletter 87, 11

\reference{} Glazebrook K., Bland-Hawthorn J., Farrell T. J., Waller L.,
Barton J. R., Lewis I. J.,  1999, AAO Newsletter, 90, 11

\reference{} Glazebrook, K. 2000, In Encyclopaedia of Astronomy and Astrophysics,
(MacMillan and Inst. of Physics Publ.,) ed. P.G. Murdin

\reference{} Glazebrook, K.\etal\ 2000a, in preparation

\reference{} Glazebrook, K.\etal\ 2000b, in preparation

\reference{} Hogg D. W., Pahre M. A., Mccarthy J. K., Cohen J. G., Blandford R., 
Smail I., Soifer B. T., 1997, MNRAS, 288, 404

\reference{} Janesick, J. \& Elliott, T. 1992, In Astronomical CCD Observing and
Reduction Techniques, ed. S.B. Howell, ASP Conf., 23, 1

\reference{} Kondratyev, K. Ya. 1969, Radiation in the Atmosphere, (New York: Acad. Press)

\reference{} Kozlowski L.J., 1996, SPIE, 2745, 2

\reference{} Leinert, Ch., Bowyer, S., Haikala, L. K., Hanner, M. S., Hauser, M. G.,
  Levasseur-Regourd, A.-Ch., Mann, I., Mattila, K., Reach, W. T.,
  Schlosser, W., Staude, H. J., Toller, G. N., Weiland, J. L.,
  Weinberg, J. L., Witt, A. N., 1998, A\&AS, 127, 1

\reference{} Lemonier, M. \& Piaget, C. 1983, IEEE Trans. Elec. Dev., 30, 1414

\reference{} McCracken H. J., Metcalfe N., Shanks T., Campos A., Gardner J. P., 
Fong R., 2000, MNRAS, 311, 707

\reference{} McLean, I.S., Cormack, W.A., Herd, J.T. \& Aspin, C. 1981, 
In Solid State Imagers for Astronomy, ed. J.C. Geary and D.W. Latham, Proc. 
SPIE, 290, 155

\reference{} McLean, I.S. 1997, Electronic Imaging in Astronomy: Detectors and 
Instrumentation (Wiley-Praxis Series in Astronomy and Astrophysics) 

\reference{} Rieke, G.H. 1994, Detection of Light: from the Ultraviolet to 
the Submillimetre, (Cambridge University Press).

\reference{} Ramsay, S.K., Mountain, C.M. \& Geballe, T.R. 1992, MNRAS, 259, 751

\reference{} Sembach K.R., Tonry J.L., 1996, AJ, 112, 797 

\reference{} Smail, I., Hogg, D.W., Lin, Y., \& Cohen, J.G. 1995, ApJ, 449, L105

\reference{} Stockman, H.S. 1982, In Instrumentation in Astronomy IV,  ed. D.L. 
Crawford, Proc. SPIE 331, 76

\reference{} Szeto, K., Stilburn J., Bond T., Roberts S., Sebesta J., 
Saddlemyer L., 1996, In Optical Telescopes of Today and Tomorrow, 
SPIE, 2871, 1262

\reference{} Wynne, C.G. \& Worswick, S.P., 1988, The Observatory, 108, 161

\reference{} Wyse R. F. G., Gilmore G., 1992, MNRAS, 257, 1


\end{references}
\end{document}